%% file: main.tex
\documentclass[a4paper, amsfonts, amssymb, amsmath, reprint, showkeys, nofootinbib, twoside,superscriptaddress]{revtex4-1}
\usepackage[english]{babel}
\usepackage[utf8]{inputenc}
\input{preamble}
\usepackage{mathrsfs,amsmath} 
\usepackage{amsfonts}
\usepackage{amssymb}
\usepackage{float}

\usepackage{caption}
\usepackage{adjustbox}

\usepackage{verbatim}
\usepackage[pdftex, pdftitle={Article}, pdfauthor={Author}]{hyperref} 

\allowdisplaybreaks
\begin{document}

\title{Stationary Distributions of the Mode-switching Chiarella Model}

\author{Jutta G. Kurth}
\affiliation{EconophysiX Lab, Institut Louis Bachelier, 28 Pl. de la Bourse, Palais Brongniart,
75002 Paris, France}
\affiliation{LadHyX UMR CNRS 7646, \'Ecole polytechnique, 91128 Palaiseau Cedex, France}

\author{Jean-Philippe Bouchaud}
\affiliation{Capital Fund Management, 23 Rue de l’Universit\'e, 75007 Paris, France}
\affiliation{EconophysiX Lab, Institut Louis Bachelier, 28 Pl. de la Bourse, Palais Brongniart,
75002 Paris, France}
\affiliation{Académie des Sciences, 23 Quai de Conti, 75006 Paris, France\smallskip}

\date{\today} 

\begin{abstract}
We derive stationary distributions of the so-called mispricing and the trend signal in various regimes of the extended Chiarella model of financial markets. This model is a stochastic nonlinear dynamical system that encompasses dynamical competition between a (saturating) trending and a mean-reverting component. We find the so-called mispricing distribution and the trend distribution to be unimodal Gaussians in the small noise, small feedback limit. Slow trends yield Gaussian-cosh mispricing distributions that allow for a P-bifurcation: unimodality occurs when mean-reversion is fast, bimodality when it is slow. The critical point of this bifurcation is established and refutes previous ad-hoc reports and differs from the bifurcation condition of the dynamical system itself. For fast, weakly coupled trends, deploying the Furutsu-Novikov theorem reveals that the result is again unimodal Gaussian. For the same case with higher coupling we disprove another claim from the literature: bimodal trend distributions do not generally imply bimodal mispricing distributions. The latter becomes bimodal only for stronger trend feedback. The exact solution in this last regime remains unfortunately beyond our proficiency. 
\end{abstract}


\maketitle

\section{Introduction}\label{sec:intro}

The Chiarella model is a nonlinear, stochastic dynamical system encompassing both negative (mean-reversion) and positive (trend following) feedback loops \cite{chiarella1992dynamics}. It was introduced in the context of financial markets to describe the dynamical interplay between value investors and trend followers. It is indeed empirically well established that (normalised) price increments, a.k.a. returns, are positively auto-correlated on short to medium time scales (weeks up to several months) -- observable as financial bubbles or trends -- while they are negatively auto-correlated on longer times scales (months to few years) -- observable as price mean reversion or corrections -- see e.g. the discussion and references in \cite{bouchaud2017black, majewski2020co}.

The model was later extended to allow for a time dependent fundamental value, which is the dynamic mean-reversion level around which the price is anchored. This level is modeled as a drift-diffusion process, which may be regarded as the fair or rationally justifiable price according to, e.g., company fundamentals in the case of stocks, or other economic indicators for other asset classes, such as indices, bonds, or derivatives \cite{majewski2020co}. Such a fundamental value only changes because of unpredictable news or ``shocks''. Devotees of the Efficient Market Hypothesis (EMH) believe that prices usually reflect all publicly available information, and that this information is instantaneously digested by the capital markets, suggesting that price and fundamental value should usually be in very close proximity. If this were true, returns should be serially uncorrelated and not exhibit the complex auto-correlation structures mentioned above, which, as many believe, heavily damage the credibility of the EMH for all major asset classes.

In a very recent paper, we amended some analytical shortcomings of the modified Chiarella model proposed in \cite{majewski2020co}. In particular, we allowed for an arbitrary time-dependent drift for the value process \cite{kurth2025revisiting}. The model was deployed and calibrated on individual assets' (log-)prices belonging to four different asset classes. What all these variations around the initial Chiarella model have in common is the existence of two distinct dynamical behaviours (in the absence of noise): 
\begin{enumerate}
    \item Attraction/convergence of price towards fundamental value;\\
    \item Oscillation of price around the fundamental value.
\end{enumerate}
It is common belief that in the presence of noise the distribution of {\it{mispricings}} (i.e. the difference between price and value) is unimodal in case 1 and bimodal in case 2, when the price stochastically quasi-oscillates around the fundamental value \cite{chiarella2011stoch_bif,majewski2020co, chiarella2008stochastic_bifurc}. This means that the phenomenological P-bifurcation condition -- dictating uni- vs. bimodality -- should coincide with the bifurcation condition predicting the transition from convergence to oscillation. While this has been argued in \cite{chiarella2011stoch_bif} and is correct in some limiting cases, the present paper disproves the result in general and provides the correct stationary mispricing distributions in many possible scenarios. The dynamical mechanisms that lead to either uni- or bimodality are clearly established.

Several extensions to the Chiarella model and other financial agent-based models have been discussed; see, e.g., \cite{majewski2020co, Goldman1980, lux1998, lux1999scaling} and references therein. A first numerical study of the stationary measure in the Chiarella model was carried out in \cite{chiarella2008stochastic_bifurc} and later in \cite{chiarella2011stoch_bif}.

Bimodalities in empirical mispricing distributions of assets in the S\&P~500 have been reported in Ref.~\cite{schmitt2017bimodality}, emphasising the real-world significance of our work, where further a number of seminal agent-based financial market models are numerically analysed with respect to their distributional mispricing bimodality. Moreover, they have been reported on spot prices of four other asset classes in Refs.~\cite{majewski2020co, kurth2025revisiting}.

This paper is organised as follows: Sec.~\ref{sec:model} introduces the Chiarella model formally and presents the crucial analytical results that are known about it. The heart of this work -- the stationary distributions of the Chiarella model in most limiting cases -- is in Secs.~\ref{sec:statdist_betagamma_small}, \ref{sec:statdist_smallalpha} and \ref{sec:alpha_large}. 
Sec.~\ref{sec:conclusion} summarises the results and concludes the main part of this paper, while additional derivations and proofs are given in Appendices.

\section{An Extended Chiarella Model}
\label{sec:model}
The Chiarella model is a stochastic dynamical system defined as
\begin{align}
\label{eq: ModifiedChiarella}
    \dd P_t &= \kappa (V_t-P_t) \dd t + \beta \tanh (\gamma M_t) \dd t + g_t \dd t + \sigma_N \dd W_t^N\nonumber \\
    \dd M_t &= -\alpha M_t \dd t+ \alpha (\dd P_t -g_t  \dd t) \\
    \dd V_t &= g_t \dd t + \sigma_V \dd W_t^V,\nonumber
\end{align}
where $\alpha$, $\kappa$, $\beta$, $\gamma$, $\sigma_{N/V}$ are all fixed positive parameters and $W^{N/V}_t$ are standard Brownian Motions. 

In the original context of the Chiarella model, $P$ is the (log-)price of a financial asset, $M$ is the trend signal, which is an exponential moving average of past drift-adjusted (log-) price increments, and $V$ is the fundamental value of the asset, modeled as a drift-diffusion process with time-dependent drift $g$. However, this model may be understood as a general dynamical system exhibiting a dynamical interplay or competition between a mean-reverting force acting on $P$, driving it towards $V$ through $\kappa (V_t-P_t)$, which is an Ornstein-Uhlenbeck (OU) component with possibly time-dependent mean-reversion level $V$, and a positive feedback term, $\beta \tanh (\gamma M)$, accounting for the trending that gives rise to temporary, larger deviations of $P$ from $V$ but that are bounded to prevent divergence or run-aways.

In order to simplify the dynamical study of the system of Eqs.\eqref{eq: ModifiedChiarella}, the model dimensionality is reduced by one by considering the mispricing amplitude $\delta := P-V$ instead of the two quantities separately without loss of generality. The model then reads
\begin{align}
\label{eq: ModifiedChiarella_delta}
    \dd \delta_t &= -\kappa \delta_t \,\dd t + \beta \tanh (\gamma M_t) \,\dd t   + \sigma_N \dd W^\textup{N}_t - \sigma_V\dd W^\textup{V}_t\nonumber \\
    \dd M_t &= -\alpha M_t \dd t+ \alpha (\dd \delta_t + \sigma_V  \dd W^\textup{V}_t).
\end{align}
Note that since $\delta$ is dimensionless, one has $[t]=[\gamma]=[T]$ and $[\alpha]=[\kappa]=[\beta]=[\sigma^2]=[M]=[T]^{-1}$. In particular, $\gamma$ describes the scale, on which the trend signal saturates, $\alpha$ is the forget rate of the exponential moving average that determines the trend $M$, $\kappa$ is the reversion speed of price to value, $\beta$ describes the typical inverse time scale of a trending phase and $\sigma_{N/V}^2$ are the typical noise frequencies.

A linear stability analysis of the deterministic counterpart ($\sigma_N=\sigma_V = 0$) of system~\eqref{eq: ModifiedChiarella_delta} reveals that it encompasses two different dynamical phases: the system undergoes a \textit{supercritical Hopf-bifurcation}, in which the loss of stability of a formerly stable fix point located at $(\delta^\star, \, M^\star) = (0, \,0)$ when $\alpha ( 1 - \beta \gamma) + \kappa > 0$ coincides with the emergence of a stable limit cycle in the $\delta$-$M$-plane when $\alpha ( 1 - \beta \gamma) + \kappa < 0$. This means that the deterministic $P$ no longer converges to the deterministic $V$ but moves around it periodically \cite{kurth2025revisiting}.
Note that both $\underset{{t\to\infty}}{\lim}\mathbb{E} [\delta]=0$ and $\underset{{t\to\infty}}{\lim}\mathbb{E} [M]=0$ whenever a stationary distribution $p(\delta, M)$ exists, since both variables obey mean-reversion forces pulling them towards zero, the hyperbolic tangent is symmetric around zero and the noises are unbiased. This can also be seen by considering that the trajectories are either spiraling into the (stable) fixed point $(\delta^\star, \, M^\star) = (0, \,0)$ or oscillating around it.

In the following sections, the stationary distributions of the system of Eqs.~\eqref{eq: ModifiedChiarella_delta} will be derived in different parameter limits using different Fokker-Planck equation (FPE) {\it ans\"atze}.

\section{The linear regime $\gamma \to 0$}
\label{sec:statdist_betagamma_small}
In the limit where $\gamma\to 0$, the hyperbolic tangent may be linearised, corresponding to its first-order Taylor expansion. The region of validity of such an expansion will be determined {\it a posteriori} as $\gamma \sigma_M \ll 1$, where $\sigma_M$ is given by Eq. \eqref{eq:sigmaM} below. Physically, this means that the demand does not saturate for the given trends $M$.

In matrix-form and defining $\mathbf{x}_t := (\delta_t, M_t)^\textup{T}$ and $\dd \mathbf{W}_t := (\dd W^\textup{N}_t, \, \dd W^\textup{V}_t)^\textup{T}$, the dynamics reads
\begin{equation}
    \dd \mathbf{x}_t = \mathbf{A}\mathbf{x}_t \dd t + \mathbf{B} \,\dd \mathbf{W}_t,
\end{equation}
where
\begin{equation}
    \mathbf{A} = \begin{pmatrix}
        -\kappa & \beta\gamma \\
        -\alpha\kappa & \alpha (\beta\gamma-1)
    \end{pmatrix}
\end{equation}
is the drift matrix of the linearised system and
\begin{equation}
    \mathbf{D}=  \mathbf{B} \mathbf{B}^\textup{T}=
    \begin{pmatrix}
        \sigma^2 & \alpha \sigma_N^2 \\
        \alpha \sigma_N^2 & \alpha^2 \sigma_N^2
    \end{pmatrix} \,\,\text{with}\,\,\,
    \mathbf{B} = \begin{pmatrix}
        \sigma_N & -\sigma_V \\ \alpha \sigma_N & 0
    \end{pmatrix}
\end{equation}
is the diffusion matrix with $\sigma = \sqrt{\sigma_N^2 + \sigma_V^2}$.

The evolution of the joint probability density $p(\delta, M, t)$ in time $t$ and the two space variables $\delta$ and $M$ for such an Ito-process is given by its corresponding FPE. The stationary distribution $p(\delta, M)$, for which $\frac{\partial p}{\partial t}=0$, is defined via the stationary FPE,
\begin{multline}\label{eq:FPE_linearised}
    0  = -\frac{\partial}{\partial \delta} \left[ (-\kappa \delta + \beta \gamma M) p \right] - \alpha
    \frac{\partial}{\partial M} \left[ (- M - \kappa\delta + \beta \gamma M) p \right] \\
    + \frac{\sigma^2 }{2} \frac{\partial^2 p}{\partial \delta^2} + \alpha\sigma_N^2 \frac{\partial^2 p}{\partial\delta\partial M } + \frac{\alpha^2 \sigma_N^2 }{2} \frac{\partial^2 p}{\partial M^2},
\end{multline}
where the left hand side of the equation is the time derivative equating zero.

Since the dynamical system is linear and its noise additive, the solution to this FPE is a (bivariate) Gaussian distribution,
\begin{equation}
p(\delta, M) = \frac{1}{2\pi \sqrt{|\mathbf{\Sigma}|}} \exp\left(-\frac{1}{2} \mathbf{x}^T \mathbf{\Sigma}^{-1} \mathbf{x}\right),
\end{equation}
where the covariance matrix $\mathbf{\Sigma}$ is
\begin{equation}
\mathbf{\Sigma} = \begin{pmatrix} \sigma_\delta^2 & \rho \sigma_\delta \sigma_M \\ \rho \sigma_\delta \sigma_M & \sigma_M^2 \end{pmatrix};
\end{equation}
$\sigma_\delta^2$ and $\sigma_M^2$ are the variances of $\delta$ and $M$ and $\rho$ is their correlation coefficient. The stationary distribution $p(\delta, M)$ is centered in zero because both $\delta$ and $M$ have vanishing mean in the stationary limit as discussed in Sec.~\ref{sec:model}.

The covariance matrix $\mathbf{\Sigma}$ in the stationary limit can be determined via the Lyapunov equation \cite{vankampen1992stochastic}
\begin{equation}
    \mathbf{A}\mathbf{\Sigma} +  \mathbf{\Sigma}\mathbf{A}^\textup{T} + \mathbf{D} = 0.
\end{equation}
The solution to this linear system of equations yields the components of $\mathbf{\Sigma}$ (see also Appendix~\ref{app:proof_statdist_betagamma_small}):
\begin{align} \label{eq:sigmadelta}
    &\sigma_\delta^2 = \frac{(\kappa + \alpha (\beta\gamma -1)^2)\sigma^2 + \alpha\beta\gamma (2-\beta\gamma)\sigma_N^2}{2\kappa (\alpha (1-\beta\gamma)+\kappa)},\\ \label{eq:sigmaM}
    &\sigma_M^2 =\frac{\alpha(\kappa \sigma^2 + (\alpha-\kappa)\sigma_N^2)}{2(\alpha (1-\beta\gamma)+\kappa)}, 
\end{align}
\begin{widetext}
\begin{equation}
\rho = \frac{\sqrt{\alpha\kappa} ((\beta\gamma-1)\sigma^2 + (2-\beta\gamma)\sigma_N^2)}{\sqrt{[(\alpha (\beta\gamma-1)^2+\kappa)\sigma^2 + \alpha\beta\gamma (2-\beta\gamma)\sigma_N^2][\kappa\sigma^2 + (\alpha-\kappa)\sigma_N^2]}}.
\end{equation}
\end{widetext}
$\mathbf{\Sigma}$ is positive semi-definite when $\kappa>\alpha(\beta\gamma-1)$, which is the bifurcation condition in the deterministic system (comp. Sec.~\ref{sec:model}). This is always true in the considered limit. If it was not true, the drift matrix $\mathbf{A}$ would have positive eigenvalues, i.e. the system would diverge and no stationary distribution would exist.
From the joint probability distribution $p(\delta, M)$, the mispricing distribution $p(\delta)$ can be obtained through marginalisation,
\begin{align}
    \label{eq: statdist_betagamma_small}
    p(\delta) = \int_{-\infty}^\infty p(\delta, M) \, \mathrm{d}M = \frac{1}{\sqrt{2\pi \sigma_\delta^2}} \mathrm{e}^{-\frac{\delta^2}{2\sigma_\delta^2}},
\end{align}
such that in this limit the mispricing distribution is Gaussian and thus unimodal. The stationary distribution of the trend signal $M$ is Gaussian, too, and can be obtained analogously. As anticipated above, these results hold provided the condition $\gamma \sigma_M \ll 1$ is satisfied. For $\alpha \gg \kappa$, this condition simplifies to
\[
\frac{\gamma^2 \alpha \sigma_N^2}{2(1 - \beta \gamma)} \ll 1,
\]
which breaks down as $\beta \gamma \to 1$. As we shall see later, this is indeed the condition for bimodality when $\alpha \gg \kappa$. 

\begin{figure}[htbp]
     \includegraphics[width=0.85\linewidth]{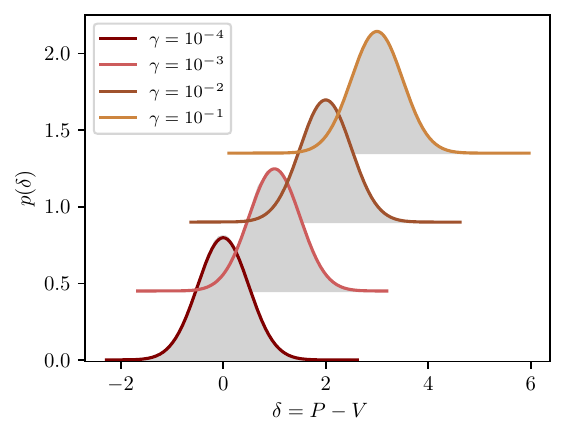}
    \caption{Grey: Numerical histograms of the stationary distribution in the case $\gamma$ small. Simulation parameters are $(\kappa,\,\beta,\,\alpha,\,\sigma_N,\,\sigma_V)=(0.1,\,0.2,\,0.2,\,0.2,\,0.1)$ and $\gamma$ as detailed in the plot. $T=\gamma\times 10^9$, d$t=\gamma / 2$ and $g=0$. Coloured: Corresponding analytical stationary distributions $p(\delta)$ according to Eq.\eqref{eq: statdist_betagamma_small}. The distributions with $\gamma >10^{-4}$ are shifted by multiples of 1 on the abscissa and of 0.5 on the ordinate.}
    \label{fig:stat_dist_betagammasmall}
\end{figure}

The numerical confirmation of this result is provided in Fig.~\ref{fig:stat_dist_betagammasmall}, which shows the numerically obtained distribution $p(\delta)$ (grey histogram) alongside the analytically derived distribution (coloured curves) for four orders of magnitude of $\gamma$. The stochastic integration (as well as all subsequent ones) was performed using the Euler-Maruyama scheme.

We now turn to two solvable limits, where either the dynamics of $\delta$ is much faster than that of $M$ ($\kappa \gg \alpha$), or vice-versa.

\section{Slow Trends: the $\kappa \gg \alpha$ Limit}
\label{sec:statdist_smallalpha}
A change of variables $x=\delta$ and $y=M-\alpha\delta$ (comp. Appendix~\ref{app:change_of_variables}) yields the following rephrasing of Eqs.~\eqref{eq: ModifiedChiarella_delta} (as before: $\sigma^2 = {\sigma_N^2 + \sigma_V^2}$):
\begin{align}
    \dd x &= -\kappa x\,\dd t + \beta \tanh (\gamma (y+\alpha x))\,\dd t + \sigma \, \dd W_t \nonumber\\
    \dd y &= -\alpha y \,\dd t + \alpha^2\, \dd x + \alpha \sigma_V \,\dd W_t^V.
    \label{eq:2Dmodel_changeVariables}
\end{align}
Since $x$ is much faster than $y$ when $\alpha\ll\kappa$, one can approximate $y$ as an OU-process, whose stationary distribution is known to be Gaussian, i.e.
\begin{equation}
    p(y) = \frac{1}{\sqrt{2\pi \mathrm{Var}[y]}} \mathrm{e}^{-\frac{y^2}{2\mathrm{Var}[y]}},
\end{equation}
with, to first order in $\alpha$, $\mathrm{Var}[y]\approx \frac{\alpha}{2} \sigma_V^2$; the exact $\mathrm{Var}[y]$ will cancel out from the expression of $p(x)$. 

The conditional FPE of the dynamics of $x$ given $y$ reads:
\begin{align} \nonumber
    \frac{\partial p(x|y)}{\partial t} =& - \frac{\partial}{\partial x} \big( [-\kappa x + \beta \tanh (\gamma (y + \alpha x))]p(x|y) \big) \\ &+ \frac{\sigma^2}{2} \frac{\partial^2}{\partial x^2} p(x|y).
\end{align}

\subsubsection{\textbf{The quasi-static equilibrium}}

In the case where $x$ evolves much faster than $y$, a quasi-static approximation can be assumed, whereby the standard Maxwell-Boltzmann equilibrium is reached before $y$ has had time to vary much, i.e.
\begin{equation}
    p(x|y) = \frac{1}{A(y)} \mathrm{e}^{-\frac{\kappa x^2}{\sigma^2}} \cosh^n (\gamma (\alpha x + y)),
\end{equation}
with $n:={\frac{2\beta}{\alpha\gamma\sigma^2}}$ and the normalisation function $A(y)$ is
\begin{equation}
\label{eq:norm_A(y)}
    A(y) = \int_{-\infty}^\infty \mathrm{e}^{-\frac{1}{\sigma^2} \kappa x^2}  \cosh^n (\gamma (\alpha x + y)) \,\dd x
\end{equation}
This integral can only be calculated explicitly for integer exponents $n \in \mathbb{N}$. In this case, and defining an $\epsilon_n$ that is zero when $n$ is even and one when $n$ is odd, one finds using the Binomial Theorem that
{\small
\begin{multline}
\label{eq:Norm_A(y)_int}
    A(y) = \sqrt{\frac{\sigma^2 \pi}{\kappa}}\frac{1}{2^{n-1}} \Bigg[
\binom{n}{\frac{n - \epsilon_n}{2}} 
\cdot 
\begin{cases}
\frac{1}{2} & n \text{ even} \\
\cosh(\gamma y) \cdot \mathrm{e}^{\frac{(\alpha\gamma\sigma)^2}{4\kappa}} & n \text{ odd}
\end{cases} \\
+ \sum_{j = 1}^{\lfloor n/2 \rfloor} 
\binom{n}{\frac{n - \epsilon_n}{2} - j} 
\cosh\left(\gamma (2j + \epsilon_n)y\right)
\mathrm{e}^{\frac{(\alpha\gamma\sigma)^2}{4\kappa } (2j + \epsilon_n)^2}
\Bigg];
\end{multline}
}
see also Appendix~\ref{app:Normal_small_alpha}.
Knowing $A(y)$, the stationary distribution $p(x)$ can in principle be calculated as
\begin{equation}
    p(x) =  \int_{-\infty}^\infty p(x|y) p(y) \,\dd y;
\end{equation}
but the solution to this integral is not known for integer exponents $n>1$ of the hyperbolic cosine.
The integral can however be solved when $\gamma$ is sufficiently large. 

\begin{figure*}
    \includegraphics[width=0.74\textwidth]{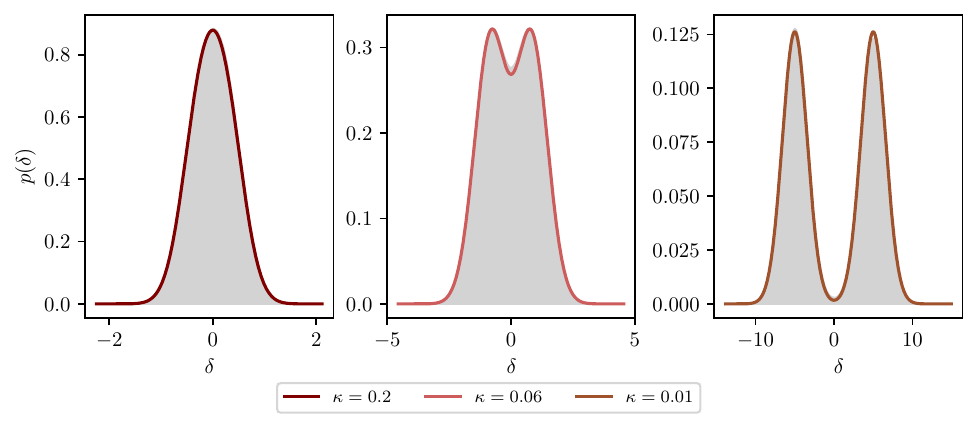}
    \caption{Same as Fig.~\ref{fig:stat_dist_betagammasmall} but for $\alpha\ll\kappa$, while $\gamma$ large, with $p(\delta)$ according to Eq.\eqref{eq:stattdist_largegamma}. Numerical parameters are $(\alpha,\,\beta,\,\gamma,\,\sigma_N,\,\sigma_V)=(2\times 10^{-5},\, 0.05,\, 5\times 10^{4},\,0.2,\,0.1)$ and $\kappa$ detailed in the plot. $T=5\times 10^7$, d$t=0.01$.}
    \label{fig:stat_dist_alphasmall2}
\end{figure*}

\subsubsection{\textbf{Large-$\gamma$ limit}}
\label{sec:small_alpha_big_gamma}
In the limit $\gamma\to\infty$, the leading exponential order of the cosh overwhelms all others, meaning that $\cosh^n (\gamma(\alpha x + y)) \approx \cosh (n\gamma(\alpha x + y))$, such that the integrands can be simplified. In this limit one does not have to assume an integer $n$. For the normalisation $A(y)$ this means
\begin{align}
    A(y) &\approx \int_{-\infty}^\infty \mathrm{e}^{-\frac{\kappa x^2}{\sigma^2}} \left( \cosh (n \gamma (\alpha x + y)) \right) \,\dd x \nonumber\\
    &= \sqrt{\frac{\pi \sigma^2}{\kappa}} \mathrm{e}^{\frac{n^2}{4\kappa} (\alpha\gamma \sigma)^2} \cosh (n\gamma y).
\end{align}
Therewith and reinserting $n=\frac{2\beta}{\alpha\gamma\sigma^2}$, the stationary mispricing distribution can be inferred:
\begin{align}
     p(\delta) 
     = p(x) &=  \int_{-\infty}^\infty p(x|y) p(y) \,\dd y \nonumber\\
    &= \sqrt{\frac{\kappa}{\pi\sigma^2}} \mathrm{e}^{-\frac{\beta^2}{\kappa \sigma^2}} \cosh \left(\frac{2\beta}{\sigma^2} \delta\right) \mathrm{e}^{-\frac{\kappa \delta^2}{\sigma^2}},
    \label{eq:stattdist_largegamma}
\end{align}
which is independent of both $\alpha$, $\gamma$; see also Appendix~\ref{app:stat_dist_small_alpha_large_gamma}. This is the Gaussian-cosh distribution that will again show up in the case $\alpha \gg \kappa$, see Appendix~\ref{app:alpha_beta_large}, Eq. \eqref{eq:PM}. The (higher-order) Gaussian-cosh distribution has previously been discussed in the context of Gaussian-cosh beam propagation in optical systems \cite{zhou2011cosh-Gaussian}.

\subsubsection{\textbf{Uni- or bimodality}}

The stationary distribution in the large-$\gamma$ limit, Eq.\eqref{eq:stattdist_largegamma}, has an extremum at $x=\delta=0$, which is either a unique maximum (unimodality)  or a minimum accompanied by two maxima symmetrically placed around it (bimodality) at solutions to $\tanh(2\beta x/\sigma^2)=\kappa x/\beta$ as $p$ is even.
The modality-type can be investigated through the curvature at $x=0$: $p''(0)\leq 0$ implies unimodality and $p''(0)>0$ bimodality, which can be summarised as
\begin{equation}
\begin{aligned}
    &\bullet \text{unimodality:} \quad \kappa \geq \frac{2\beta^2}{\sigma^2}\\
    &\bullet \text{bimodality:} \quad\,\,\,\, \kappa < \frac{2\beta^2}{\sigma^2}.
\end{aligned}
\label{eq:modality_condition_alphasmall}
\end{equation}
This result should be compared with the condition $\kappa = \alpha (\beta \gamma -1)$ for the loss of stability of the fixed point $(\delta^\star, M^\star)=(0,0)$ mentioned above. Hence we refute the claim in \cite{chiarella2011stoch_bif} that the P-bifurcation condition -- the bifurcation in the modality of the stationary distribution -- coincides with the bifurcation condition on the possible types of solutions to Eqs.~\eqref{eq: ModifiedChiarella_delta} (convergence vs. oscillation). Instead, we find that the criterion is more subtle and depends on both the noise of $P$ and $V$ through $\sigma$. Naturally, we confirm that the mean-reversion force $\sim\kappa$ works against bimodality and the trend component $\sim\beta$ induces bimodality -- albeit in a non-trivial quadratic way and not linearly as stated in \cite{chiarella2011stoch_bif}. Finally, it shows that strong enough noise can wipe out any bimodality, an intuitive result indeed.

Our claim is numerically confirmed in Fig.~\ref{fig:stat_dist_alphasmall2}, which displays several distributions with the correct number of modes according to Eqs.~\eqref{eq:modality_condition_alphasmall}. The center case of Fig.~\ref{fig:stat_dist_alphasmall2} is a case where the condition by Chiarella et al. \cite{chiarella2011stoch_bif} would have falsely predicted $p(\delta)$ to be unimodal via the Hopf-bifurcation condition of the dynamical system, while Eq.\eqref{eq:modality_condition_alphasmall} correctly predicts the bimodality, which has further been confirmed on edge cases.

\begin{figure}
    \centering
    \includegraphics[width=0.85\linewidth]{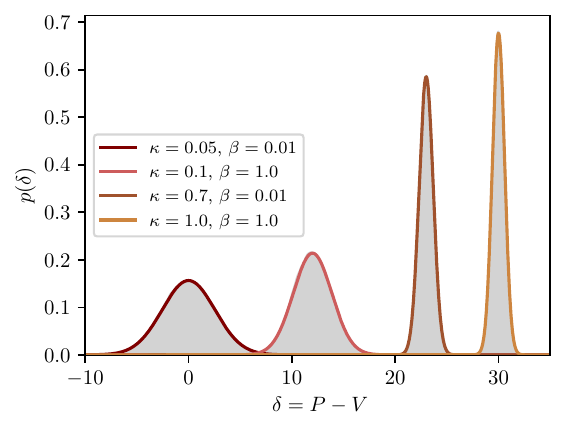}
    \caption{Same as Fig.~\ref{fig:stat_dist_betagammasmall} but in the case $\alpha\gg\kappa$ with $p(\delta)$ according to Eq.\eqref{eq:stat_dist_alpha_large}. Numerical parameters are $(\alpha,\,\gamma,\,\sigma_N,\,\sigma_V)=(500,\,1,\,0.8,\,0.1)$, while $\kappa$ and $\beta$ are detailed in the plot. $T=10^5$, d$t=10^{-3}$.}
    \label{fig:stat_dist_alpha_large}
\end{figure}

\section{Fast Trends: the $\alpha\gg\kappa$ Limit}
\label{sec:alpha_large}

Considering the Langevin equation corresponding to Eqs.~\eqref{eq:2Dmodel_changeVariables}, we see that $y$ tracks $x$ closely when $\alpha\gg\kappa$. In this case, and assuming $\gamma\sigma_N^2 \nrightarrow 0$, the hyperbolic tangent acts as an auto-correlated telegraphic noise $\xi^\textup{tele}_t\in \{\pm 1\}$. In that case first $y$ and then $x$ can be integrated out, such that the second moment $\langle x^2 \rangle$ may be derived (see Appendices~\ref{app:change_of_variables}, \ref{app:alpha_large}). It turns out that sub-cases are determined by the dimensionless parameter $\Theta:=\beta/(\sigma_N \sqrt{\alpha})$.

\subsubsection{{\bf Weak coupling:} $\Theta \ll 1$}
\label{sec:alpha_larger_kappa_beta}
The switching rate of the telegraphic noise $\xi^\textup{tele}$ is proportional to $\alpha$ in this case. When $\alpha\gg\kappa$, the distribution is then a unimodal Gaussian because the telegraphic noise switches sign much faster than any relaxation to a potential steady state at $x\approx \pm {\beta_\text{eff\,}}/{\kappa_\text{eff\,}}$ could take place (on a typical time scale $1/\kappa_\text{eff\,}$), where
\begin{equation}
    \kappa_\text{eff\,} = \kappa \left( 1 + \frac{2 \Theta}{\sqrt{\pi}} \right)\quad\text{and}\quad \beta_\text{eff\,} = \beta \left( 1 + \frac{2 \Theta}{\sqrt{\pi}}\right);
\end{equation}
see also Appendix~\ref{app:alpha_large}. The distribution in this limit reads
\begin{equation}
\label{eq:stat_dist_alpha_large}
    p(\delta) = p(x) = \mathcal{N} (0, \langle x^2 \rangle),
\end{equation}
where $\mathcal{N}$ refers to the Gaussian distribution; it has mean zero and variance
\begin{equation}
    \langle x^2 \rangle = 
    \frac{\sigma^2}{2\kappa_\text{eff\,}} +  \frac{2 \alpha \sigma_N^2}{\sqrt{\pi}\kappa (\alpha + \kappa_\text{eff\,})} \Theta +  \frac{\ln (2) \sigma_N^2}{\kappa} \Theta^2 + \mathcal{O} (\Theta^3).
\end{equation}
This is fully derived in Appendix~\ref{app:alpha_large} using the Furutsu-Novikov theorem \cite{ishimaru1978wave} and illustrated for different values of $\beta$ and $\kappa$ in Fig.~\ref{fig:stat_dist_alpha_large}.

Chiarella et al. derive an analytical distribution of the trend signal $p(M)$ for $\alpha\gg\kappa$ but only in the absence of noise traders ($\sigma_N =0$). In this case Eq.~\eqref{eq:stat_dist_alpha_large} does not hold as the condition $\gamma \sigma_N^2\nrightarrow 0$ is violated. They do not offer an analytical result for $p(\delta)$ \cite{chiarella2011stoch_bif}.

\subsubsection{{\bf Moderate to strong coupling:} $\Theta \sim 1$}

In the limit where $\alpha \gg \kappa$, $\Theta \gtrsim 1$, we can disprove another claim from the literature (comp. \cite{chiarella2011stoch_bif, majewski2020co, kurth2025revisiting}), which is that {\it bimodal trend distributions must co-occur with bimodal mispricing distributions}. This claim only holds true in the absence of noise traders ($\sigma_N =0$), which is the case regarded in \cite{chiarella2008stochastic_bifurc}. But in the presence of noise traders (which is the realistic case in the context of most physical systems, including financial markets, and our model), this result crumbles down. 

Instead, we show that when $\alpha \gg \kappa$, $\Theta \gtrsim 1$, a bimodal trend distribution $p(M)$ (again of the Gaussian-cosh type) does not imply a bimodal mispricing distribution $p(\delta)$. From Eq. \eqref{eq:PM} in Appendix~\ref{app:alpha_beta_large}, our scenario is that as $\Theta$ increases, $p(M)$ becomes bimodal while $p(\delta)$ remains unimodal before both become bimodal for $\Theta > \Theta_c$ (and thus $\beta$, the strength of the trend feedback, large enough). 

\begin{figure}[htb]
    \centering
    \includegraphics[width=0.5\textwidth]{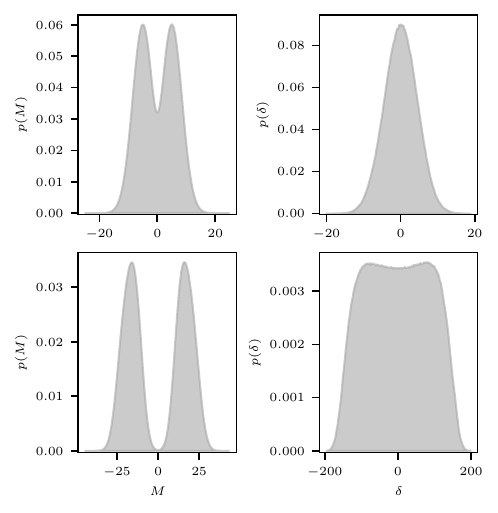}
    \caption{Same as Fig.~\ref{fig:stat_dist_betagammasmall} but for $\alpha,\,\beta\gg\kappa$, and for both $p(M)$ (left) and $p(\delta)$ (right). Top: Numerical parameters are $(\alpha,\,\kappa,\,\gamma,\,\sigma_N,\,\sigma_V)=(50,\, 0.05, 1,\,0.7,\,0.2)$, $\beta = 5$ and drift $g=0$, corresponding to $\Theta \approx 1.01$ and $\gamma \sigma_N \sqrt{\alpha} \approx 5$. $T=10^5$, d$t=0.001$. We clearly see that for this set of parameters $p(M)$ is bimodal while $p(\delta)$ is still unimodal. Bottom: same parameters as before except $\beta = 18$, such that $\Theta \approx 3.64$, in which case both distributions are bimodal. We have no exact analytical predictions to compare with in these cases.}
    \label{fig:stat_dist_alpha_beta_large}
\end{figure}

This scenario is supported by Fig.~\ref{fig:stat_dist_alpha_beta_large} (top row), which shows that there indeed exists a parameter range, in which the stationary trend distribution $p(M)$ is bimodal, while the mispricing distribution $p(\delta)$ is still unimodal; $p(\delta)$ also becomes bimodal when $\beta$ is further increased as displayed in Fig.~\ref{fig:stat_dist_alpha_beta_large} (bottom row). The underlying mechanism, detailed in Appendix~\ref{app:alpha_beta_large}, is that for intermediate values of $\beta \gtrsim 1/\gamma$ the distribution $p(M)$ becomes bimodal (see Eq.~\eqref{eq:PM}) but $M$ remains fast compared to $\delta$, so that the telegraphic noise $\xi^\textup{tele}$ cannot ``polarize'' $\delta$ long enough for $p(\delta)$ to become bimodal. When $\beta$ increases further, the dynamics of $M$ abruptly slows down -- in fact as $\alpha^{-1} e^{\Theta^2}$ --, so that $\beta \tanh(\gamma M)$ pushes $\delta$ up or down for long enough to make $p(\delta)$ bimodal. But, by the same token, the equilibration of $M$ at fixed $\delta$ ceases to be fast when $\Theta$ increases, even when $\alpha$ is large, and therefore one can no longer use the argument based on separation of time scales developed in  Appendix~\ref{app:alpha_beta_large}. The critical value of $\Theta_c \approx 0.798$ for which $p(\delta)$ becomes bimodal is thus only indicative and, as argued in Appendix~\ref{app:alpha_beta_large}, possibly a lower bound to the true value. In fact, as shown in Fig.~\ref{fig:stat_dist_alpha_beta_large}, the distribution $p(\delta)$ is still unimodal for $\Theta = 1.01$.

\section{Conclusion}\label{sec:conclusion}

In this paper, we have obtained, either exactly or approximately, the stationary distribution of an extended Chiarella model in many dynamical regimes of interest. This has led to several falsifications of results from the literature, which we have been able to amend, especially regarding the classification of the stationary distribution by its number of modes.

First, it was found that the stationary distribution is always unimodal when the dynamical system and its corresponding Fokker-Planck equation can be linearised, leading to a Gaussian stationary distribution of both the trend signal, and the mispricing, as well as their joint distribution, which is bivariate Gaussian. The condition for this to hold was explicitly computed. 

Second, the claim that the P-bifurcation (bimodality) condition coincides with the Hopf-bifurcation condition of the noiseless dynamical system has been disproved, both analytically and numerically. A corrected condition for the transition from uni- to bimodality was established when the trend time scale, $\alpha^{-1}$, and the trend saturation parameter, $\gamma$, are both large. The condition $2\beta^2 > \kappa \sigma^2$ is linear in the mean-reversion strength $\kappa$ as is the Hopf-bifurcation condition but quadratic in the trend feedback parameter $\beta$, which is different from the Hopf-bifurcation condition. Furthermore, the new condition depends on the strength of both sources of noise. This may be interpreted intuitively: strong noise wipes out the bimodality by overshadowing it.

Third, a stationary distribution in the case where the trend time scale $\alpha^{-1}$ is short compared to the typical mean-reversion time $\kappa^{-1}$, while the positive feedback term $\beta$ is not very strong, is found via the Furutsu-Novikov theorem. In this case the distribution is unimodal and Gaussian, too. If, in turn, the positive feedback is strong, we find -- disproving another common claim in the literature -- that bimodality in the trend distribution does not necessarily imply bimodality in the mispricing distribution when the price variable has its own noise source. Only when the feedback parameter $\beta$ is sufficiently strong do both distributions become bimodal.

We have unfortunately not been able to find an exact analytical solution for the stationary distribution in these last cases, for reasons that we explain in the text and in the corresponding Appendix. However, it might be possible to obtain approximate solutions in these cases as well, in particular when the trend distribution is bimodal and the trend remains polarized in the same direction for a very long time. In this case, we expect the mispricing distribution to become close to the sum of two Gaussian distributions centred around $\pm \beta/\kappa$.

In order to bring the theory derived in this work even closer to empirical findings, an asymmetric model set-up could be considered because Ref.~\cite{schmitt2017bimodality} shows using data from Ref.~\cite{shiller2000irrational} that booms are stronger than busts in stock-markets, implying bimodal mispricing distributions that are right-skewed. This could, for instance, be achieved by having asymmetric reversion strengths depending on whether the asset is over-, or underpriced.

\section*{Acknowledgements}
This research was conducted within the Econophysics \& Complex Systems Research Chair under the aegis of the Fondation du Risque, the Fondation de l’Ecole polytechnique, the Ecole polytechnique, and Capital Fund Management.

\clearpage

\bibliographystyle{vancouver}
\bibliography{biblio}

\clearpage

\appendix

\small 
\onecolumngrid

\section{Proof of the Stationary Distribution in the small-$\beta\gamma$-limit}\label{app:proof_statdist_betagamma_small}
In this appendix we proof that the bivariate Gaussian stationary distribution derived using the Lyapunov equation {\it ansatz} in Sec.~\ref{sec:statdist_betagamma_small} solves the corresponding stationary Fokker-Planck equation.
The stationary distribution is defined as 
\begin{equation}
p(\delta, M) = \frac{1}{2\pi \sqrt{|\mathbf{\Sigma}|}} \exp\left(-\frac{1}{2} \mathbf{x}^T \mathbf{\Sigma}^{-1} \mathbf{x}\right), \quad \text{with}\,\, \mathbf{x} = \begin{pmatrix}
    \delta \\ M
\end{pmatrix}.
\end{equation}
where the covariance matrix $\mathbf{\Sigma}$ and its inverse are
\begin{align}\mathbf{\Sigma} &= \frac{1}{2 (\alpha (1-\beta\gamma) + \kappa)}
    \begin{pmatrix}
 \frac{\alpha \sigma^2 (\beta \gamma -1)^2 + \alpha \beta \gamma \sigma_N^2 (2-\beta \gamma) + \kappa \sigma^2}{\kappa } & \alpha (\sigma^2 (1-\beta \gamma) + \sigma_N^2 (\beta \gamma -2)) \\
 \alpha (\sigma^2 (1-\beta \gamma) + \sigma_N^2 (\beta \gamma -2)) & \alpha (\alpha \sigma_N^2 + \kappa (\sigma^2 - \sigma_N^2)) \\
\end{pmatrix},\\
    \mathbf{\Sigma}^{-1} &= \frac{2(\alpha(\beta\gamma-1)-\kappa)}{\sigma^2 \sigma_N^2 \left(-\alpha ^2 (\beta  \gamma -1)^2+2 \alpha  \kappa  (\beta  \gamma -2)+\kappa ^2\right)+\alpha  \sigma_N^4 (\beta  \gamma -2)
   (\alpha  \beta  \gamma -2 \kappa )-\kappa ^2 \sigma ^4}\quad \times \\
&\quad\quad\quad\quad\quad\quad \quad \begin{pmatrix}
 \kappa  \left(\sigma_N^2 (\alpha -\kappa )+\kappa  \sigma ^2\right) & \kappa  \left(\sigma ^2 (1-\beta  \gamma )+\sigma_N^2 (\beta  \gamma -2)\right)
   \\
 \kappa  \left(\sigma ^2 (1-\beta  \gamma )+\sigma_N^2 (\beta  \gamma -2)\right) & \frac{\sigma ^2 \left(\alpha  (\beta  \gamma -1)^2+\kappa \right)}{\alpha
   }+\beta  \gamma  \sigma_N^2 (2-\beta  \gamma ) \\
\end{pmatrix}=:
    \begin{pmatrix}
        a & b \\ b & c
    \end{pmatrix}.
\end{align}

The partial derivatives of $p$ are
\begin{align}
\partial_\delta p &= -(a\delta + bM)p, \\
\partial_M p &= -(b\delta + cM)p, \\
\partial_\delta^2 p &= \left[(a\delta + bM)^2 - a\right]p, \\
\partial_M^2 p &= \left[(b\delta + cM)^2 - c\right]p, \\
\partial_\delta \partial_M p &= \left[(a\delta + b M)(b\delta + cM) - b \right]p.
\end{align}
Next, we insert these into the following stationary FPE, which is equivalent to the linear FPE, Eq.\eqref{eq:FPE_linearised},
\begin{equation}
0 = \kappa p + \kappa \delta \partial_\delta p - \beta \gamma M \partial_\delta p + \alpha (1-\beta\gamma) p + \alpha (1-\beta\gamma)M \partial_M p + \alpha \kappa\delta \partial_M p + \frac{\sigma^2}{2} \partial_\delta^2 p + \alpha \sigma_N^2 \partial_\delta \partial_M p + \frac{\alpha^2 \sigma_N^2}{2} \partial_M^2 p
\end{equation}
and collect the terms by order.
Since the joint stationary distribution holds for all $\delta, \, M$, each term must be zero.

\textbf{Constant terms:}
\begin{equation}
\kappa + \alpha (1 - \beta \gamma ) - \frac{\sigma^2}{2}a - \alpha \sigma_N^2 b - \frac{\alpha^2 \sigma_N^2}{2}c \overset{!}{=} 0
\end{equation}

\textbf{$\delta^2$-terms:}
\begin{equation}
-\kappa a -  \alpha \kappa b  + \frac{\sigma^2}{2}a^2 + \alpha \sigma_N^2 a b + \frac{\alpha^2 \sigma_N^2}{2}b^2 \overset{!}{=} 0
\end{equation}

\textbf{$M^2$-terms:}
\begin{equation}
\beta \gamma b - \alpha (1 - \beta \gamma) c + \frac{\sigma^2}{2}b^2 + \alpha \sigma_N^2 b c + \frac{\alpha^2 \sigma_N^2}{2}c^2 \overset{!}{=} 0
\end{equation}

\textbf{$\delta M$-terms:}
\begin{equation}
- \kappa b + \beta \gamma a  - \alpha \kappa c - \alpha (1-\beta \gamma ) b + \sigma^2 a b + \alpha \sigma_N^2 (a c + b^2) + \alpha^2 \sigma_N^2 b c \overset{!}{=} 0
\end{equation}

The solution is confirmed, $p$ solves the FPE.

\section{Change of Variables}
\label{app:change_of_variables}
Let us define $x=\delta$ and $y=M-\alpha \delta=M-\alpha x$.
Then in Langevin notation ($\xi_t^{N/V}$ are standard white noises):
\begin{align}
    \dot{x} &= -\kappa x + \beta \tanh (\gamma M) + \sigma_N \xi_t^N - \sigma_V \xi_t^V \nonumber\\
    &= -\kappa x + \beta \tanh (\gamma (y + \alpha x )) + \sigma_N \xi_t^N - \sigma_V \xi_t^V
    \label{eq:change_var_xdot}
\end{align}
And
\begin{align}
    \dot{y} &= \dot{M} -\alpha \dot{x} \\
    &= -\alpha M + \alpha (\dot{x} + \sigma_V \xi_t^V) - \alpha \dot{x} \\
    &= -\alpha M + \alpha \sigma_V \xi_t^V \\
    &= -\alpha y -\alpha^2 x + \alpha \sigma_V \xi_t^V
    \label{eq:change_var_ydot}
\end{align}

So for $\alpha\ll \kappa$ this mean that $y$ is an approximate OU process and the derivation of the stationary distribution in Sec.~\ref{sec:statdist_smallalpha} holds true.

To solve for the stationary distribution in cases where $\alpha\gg\kappa$, $\dot{y}$ first has to be solved, which can be done by multiplying both side with an integrating factor $\mathrm{e}^{\alpha t}$, and yields ($t>s$)
\begin{equation}
    y(t) = \mathrm{e}^{-\alpha t} y(0) - \alpha^2 \mathrm{e}^{-\alpha t} \int_{0}^t \mathrm{e}^{\alpha s} x(s) \,\dd s + \alpha \sigma_V \mathrm{e}^{-\alpha t} \int_{0}^t \mathrm{e}^{\alpha s} \xi^V_s \,\dd s;
\end{equation}
in the stationary limit, in which the inital condition is forgotten, this becomes (using integration by parts)
\begin{align}
    y(t) &= - \alpha^2 \int_{0}^t \mathrm{e}^{-\alpha (t-s)} x(s) \,\dd s + \alpha \sigma_V \int_{0}^t \mathrm{e}^{-\alpha (t-s)} \xi^V_s \,\dd s \nonumber\\
    &= -\alpha x(t) + \alpha \int_0^t \mathrm{e}^{-\alpha (t-s)} \dot{x}(s) \,\dd s +\alpha \sigma_V \int_{0}^t \mathrm{e}^{-\alpha (t-s)} \xi^V_s \,\dd s, \nonumber\\
\end{align}
a convolution integral. Note also that the third term is the (stationary) solution to an Ornstein-Uhlenbeck (OU) process with mean reversion level zero, mean reversion strength $\alpha$ and a variance of $\alpha\sigma_V^2 / 2$.

Inserting this into $\dot{x}$ 
yields
\begin{align}
    \label{eq: change_var_xdot_withy}
    \dot{x} &= -\kappa x + \beta \tanh (\gamma (y + \alpha x)) + \sigma_N \xi_t^N - \sigma_V \xi_t^V \nonumber\\
    &= -\kappa x + \beta \tanh (\gamma (\alpha \int_0^t \mathrm{e}^{-\alpha (t-s)} \dot{x}(s) \,\dd s +\alpha \sigma_V \int_{0}^t \mathrm{e}^{-\alpha (t-s)} \xi^V_s \,\dd s)) + \sigma_N \xi_t^N - \sigma_V \xi_t^V .
\end{align}
A special case of this Langevin equation will be further analysed in Appendix~\ref{app:alpha_large}.

\section{Additional Results in the limit $\alpha\ll\kappa$}
\subsection{Normalisation Function $A(y)$}
\label{app:Normal_small_alpha}

In this section we derive the normalisation function $A(y)$ given in Sec.~\ref{sec:statdist_smallalpha}, Eq.\eqref{eq:Norm_A(y)_int}.
As detailed in the main text, Eq.\eqref{eq:norm_A(y)}, the normalisation function reads
\begin{equation}
    A(y) = \int_{-\infty}^\infty \mathrm{e}^{-\frac{1}{\sigma^2} \kappa x^2}  \cosh (\gamma (\alpha x + y)) ^n \,\dd x
\end{equation}
and can generally only be calculated for integer exponents $n= \frac{2\beta}{\alpha\gamma\sigma^2} \in \mathbb{N}$ of the hyperbolic cosine. As before: $\sigma = \sqrt{\sigma_N^2 + \sigma_V^2}$.

Using the binomial theorem and the exponential representation of the cosh, the integral can be rewritten as
\begin{align}
   & \int_{-\infty}^\infty \mathrm{e}^{-\frac{\kappa x^2}{\sigma^2} } \frac{1}{2^n} \left( \mathrm{e}^{-\gamma\alpha x} \mathrm{e}^{-\gamma y} + \mathrm{e}^{\gamma\alpha x} \mathrm{e}^{\gamma y}  \right)^n \,\dd x \\
   =& \int_{-\infty}^\infty \mathrm{e}^{-\frac{\kappa x^2}{\sigma^2} } \frac{1}{2^n} \left( \sum_{k=0}^n \binom{n}{k}  \mathrm{e}^{-\gamma\alpha x (n-k)} \mathrm{e}^{-\gamma y (n-k)} \mathrm{e}^{\gamma\alpha x k} \mathrm{e}^{\gamma y k}   \right) \,\dd x \\
   =&  \frac{1}{2^n} \sum_{k=0}^n \binom{n}{k}  \mathrm{e}^{-\gamma (n-2k)y} \int_{-\infty}^\infty \mathrm{e}^{-\frac{\kappa x^2}{\sigma^2}- \alpha\gamma  (n-2k)x} \,\dd x \\
   = &  \frac{1}{2^n} \sqrt{\frac{\pi \sigma^2}{\kappa}} \sum_{k=0}^n \binom{n}{k} \mathrm{e}^{-\gamma (n-2k) y + \frac{(\alpha\gamma\sigma)^2}{4\kappa} (n-2k)^2},
\end{align}
where in the last step the Gaussian identity $\int_{-\infty}^\infty \mathrm{e}^{-(ax^2 + b x)} = \sqrt{\frac{\pi}{a}} \mathrm{e}^{\frac{b^2}{4a}}$ is used.

Now regard
\[
\frac{1}{2^n} \sum_{k=0}^{n} \binom{n}{k} \exp\left[-\gamma (n-2k) y + \frac{(\alpha\gamma\sigma)^2}{4\kappa} (n-2k)^2\right].
\]
To simplify, a change of variables is performed. Define
\[
m_j := 2j + \epsilon_n, \quad \text{where } \epsilon_n = 
\begin{cases}
0 & \text{if } n \text{ is even} \\
1 & \text{if } n \text{ is odd.}
\end{cases}
\]
Then \(j\) runs over integers from \(-\lfloor n/2 \rfloor\) to \(\lfloor n/2 \rfloor\), where $\lfloor.\rfloor$ marks the floor-function for integer division, and the binomial coefficient becomes:
\[
\binom{n}{\frac{n - m_j}{2}} = \binom{n}{\frac{n - (2j + \epsilon_n)}{2}} = \binom{n}{\frac{n - \epsilon_n}{2} - j}
\]
The sum then reads
\[
\frac{1}{2^n} \sum_{j = -\lfloor n/2 \rfloor}^{\lfloor n/2 \rfloor} 
\binom{n}{\frac{n - \epsilon_n}{2} - j}
\exp\left[-\gamma (2j + \epsilon_n) y + \frac{(\alpha\gamma\sigma)^2}{4\kappa} (2j + \epsilon_n)^2\right]
\]
To further simplify the result, the symmetry of the sum may be exploited: The linear term in the exponent is odd in \(j\), while the quadratic term and the binomial coefficient are even in \(j\) due to the identity:
\[
\binom{n}{\frac{n - \epsilon_n}{2} - j} = \binom{n}{\frac{n - \epsilon_n}{2} + j}.
\]
Thus, the overall sum is symmetric (for $j\neq 0$), and  \(j\)- and \(-j\)-terms can be grouped together:
\[
\exp\left[-\gamma  m_j y + \frac{(\alpha\gamma\sigma)^2}{4\kappa} m_j^2\right] + 
\exp\left[+\gamma m_j y + \frac{(\alpha\gamma\sigma)^2}{4\kappa} m_j^2\right] = 
2 \cosh(\gamma m_j y) \cdot \exp\left(\frac{(\alpha\gamma\sigma)^2}{4\kappa} m_j^2\right).
\]
Therefore, the final expression reads
\begin{align*}
& \frac{1}{2^n} \bigg[
\binom{n}{\frac{n - \epsilon_n}{2}} \mathrm{e}^{\frac{(\alpha\gamma\sigma)^2}{4\kappa} \epsilon_n^2} \left[ 2\epsilon_n \cosh (\gamma y) + (1-\epsilon_n) \right]   + 2 \sum_{j = 1}^{\lfloor n/2 \rfloor} 
\binom{n}{\frac{n - \epsilon_n}{2} - j} 
\cosh\left(\gamma (2j + \epsilon_n) y \right)
\mathrm{e}^{\frac{(\alpha\gamma\sigma)^2}{4\kappa} (2j + \epsilon_n)^2}
\bigg] \\
=& \frac{1}{2^{n-1}} \left[ \frac{1}{2}
\binom{n}{\frac{n - \epsilon_n}{2}}  \left( 2\epsilon_n \cosh (\gamma y) \mathrm{e}^{\frac{(\alpha\gamma\sigma)^2}{4\kappa}} + (1-\epsilon_n) \right)
+  \sum_{j = 1}^{\lfloor n/2 \rfloor} 
\binom{n}{\frac{n - \epsilon_n}{2} - j} 
\cosh\left(\gamma (2j + \epsilon_n) y \right)
\mathrm{e}^{\frac{(\alpha\gamma\sigma)^2}{4\kappa} (2j + \epsilon_n)^2}
\right].
\end{align*}
This is the simplified and symmetrized form of the original sum, now including the binomial weights and expressed as a weighted sum of hyperbolic cosine functions. The first two terms are the case distinction for $j=0$ (then $\epsilon_n =0$ when $n$ is even but $\epsilon_n =1$ when $n$ is odd), which is a cosh for odd $n$ as the index never becomes zero in that case such that the symmetry relation still holds. For even $n$, the index goes through zero (as the decrements are in steps of 2), in which case there is only one summand and no pairing and the summand becomes $1 \sim \mathrm{e}^0$.

This concludes the proof of Eq.\eqref{eq:Norm_A(y)_int}, i.e.
\[ A(y) = 
\frac{1}{2^{n-1}} \sqrt{\frac{\pi \sigma^2}{\kappa}} \left[
\binom{n}{\frac{n - \epsilon_n}{2}} 
\cdot 
\begin{cases}
\frac{1}{2} & \text{if } n \text{ is even} \\
\cosh(\gamma y) \cdot \mathrm{e}^{\frac{(\alpha\gamma\sigma)^2}{4\kappa}} & \text{if } n \text{ is odd}
\end{cases}
+ \sum_{j = 1}^{\lfloor n/2 \rfloor} 
\binom{n}{\frac{n - \epsilon_n}{2} - j} 
\cosh\left(\gamma (2j + \epsilon_n) y \right)
\mathrm{e}^{\frac{(\alpha\gamma\sigma)^2}{4\kappa} (2j + \epsilon_n)^2}
\right].
\]

\subsection{Large-$\gamma$-Limit Derivations: Normalisation and Stationary Distribution}
\label{app:stat_dist_small_alpha_large_gamma}

In the limit $\gamma \to \infty$, while $\alpha\gamma=\text{const.} \in \mathbb{R}$, the stationary distribution can be derived. For that, $A(y)$ must first be determined. When the exponents in $A(y)$ are large, the leading order term substantially overwhelms all other terms. Therefore, taking the large-$\gamma$-limit (\textit{or} the large-$n$-limit), allows for the following approximations:
\begin{align}
     (\mathrm{e}^{\alpha\gamma x} \mathrm{e}^{\gamma y} + \mathrm{e}^{-\alpha\gamma x} \mathrm{e}^{-\gamma y})^n &\approx 
     \begin{cases}
         \mathrm{e}^{n\alpha\gamma x} \mathrm{e}^{n\gamma y},\,\, x>0 \\
         \mathrm{e}^{-n\alpha\gamma x} \mathrm{e}^{-n\gamma y},\,\, x<0
     \end{cases} \\
     &\approx
      \mathrm{e}^{n\alpha\gamma x} \mathrm{e}^{n\gamma y} + \mathrm{e}^{-n\alpha\gamma x} \mathrm{e}^{-n\gamma y} \approx 2^n \cosh (n \gamma (\alpha x + y)),
\end{align}
where we have combined the cases $x>0$ and $x<0$ as one of the terms is asymptotically zero in either case and thus does not impact the asymptotic expansion. The same result could be obtained by performing the two integrals individually and adding the results back together for the full solution.
Thus, the normalisation is the following in this limit:
\begin{align}
    A(y) &\approx \int_{-\infty}^\infty \mathrm{e}^{-\frac{\kappa x^2}{\sigma^2}} \left( \cosh (n \gamma (\alpha x + y)) \right) \,\dd x \\
    &= \frac{1}{2} \sqrt{\frac{\pi \sigma^2}{\kappa}} \mathrm{e}^{\frac{n^2}{4\kappa}(\alpha\gamma \sigma)^2} \mathrm{e}^{-n\gamma y} \left( 1 + \mathrm{e}^{2n\gamma y} \right) \\
    &= \sqrt{\frac{\pi \sigma^2}{\kappa}} \mathrm{e}^{\frac{n^2}{4\kappa} (\alpha\gamma \sigma)^2} \cosh (n\gamma y).
\end{align}
Using the normalisation and the same asymptotic results above, the stationary distribution can be calculated
\begin{align}
    p(x) &= \int_{-\infty}^\infty p(x|y) p(y) \,\dd y \\
    &= \frac{1}{\sqrt{2\pi Var[y]}} \mathrm{e}^{-\frac{\kappa x^2}{\sigma^2}}
    \int_{-\infty}^\infty \frac{1}{A(y)} \mathrm{e}^{\frac{-y^2}{2Var[y]}} \cosh (\gamma (\alpha x + y))^n \,\dd y \\
    & \approx \frac{1}{\sqrt{2\pi Var[y]}} \sqrt{\frac{\kappa}{\pi\sigma^2}} \mathrm{e}^{-\frac{n^2}{4\kappa} (\alpha\gamma \sigma)^2} \mathrm{e}^{-\frac{\kappa x^2}{\sigma^2}}
    \int_{-\infty}^\infty \frac{1}{\cosh (n\gamma y)} \mathrm{e}^{\frac{-y^2}{2Var[y]}} \cosh (n\gamma (\alpha x + y)) \,\dd y \\
    &= \frac{1}{\sqrt{2\pi Var[y]}} \sqrt{\frac{\kappa}{\pi\sigma^2}} \mathrm{e}^{-\frac{n^2}{4\kappa} (\alpha\gamma \sigma)^2} \mathrm{e}^{-\frac{\kappa x^2}{\sigma^2}}
    \int_{-\infty}^\infty \frac{\mathrm{e}^{\frac{-y^2}{2Var[y]}}}{\cosh (n\gamma y)}  \left[ \cosh (n\gamma\alpha x)\cosh (n\gamma  y) + \sinh (n\gamma\alpha x)\sinh (n\gamma  y) \right] \,\dd y \\
    &= \frac{1}{\sqrt{2\pi Var[y]}} \sqrt{\frac{\kappa}{\pi\sigma^2}} \mathrm{e}^{-\frac{n^2}{4\kappa} (\alpha\gamma \sigma)^2} \mathrm{e}^{-\frac{\kappa x^2}{\sigma^2}}
    \left[ \cosh (n\alpha\gamma x) \int_{-\infty}^\infty \mathrm{e}^{\frac{-y^2}{2Var[y]}} \, \dd y + \sinh (n\alpha\gamma x) \int_{-\infty}^\infty \mathrm{e}^{\frac{-y^2}{2Var[y]}} \tanh (n\gamma y) \,\dd y
    \right] \\
    &= \sqrt{\frac{\kappa}{\pi\sigma^2}} \mathrm{e}^{-\frac{n^2}{4\kappa} (\alpha\gamma \sigma)^2} \cosh (n \alpha\gamma x) \mathrm{e}^{-\frac{\kappa x^2}{\sigma^2}},
\end{align}
where the second integral equates zero because its integrand is odd over the symmetric integration domain.

Recalling $n=\frac{2\beta}{\alpha\gamma \sigma^2}$, the stationary distribution in the large-$\gamma$(or $n$)-limit finally reads
\begin{equation}
    p(x) = \sqrt{\frac{\kappa}{\pi\sigma^2}} \mathrm{e}^{-\frac{\beta^2}{\kappa \sigma^2}} \cosh \left(\frac{2\beta}{\sigma^2} x\right) \mathrm{e}^{-\frac{\kappa x^2}{\sigma^2}}.
\end{equation}
Two things can be observed:
\begin{enumerate}
    \item The distribution lost its $\gamma$-dependence (as expected) but this leads to the stationary distribution being independent of the trend time scale $\alpha$, too. Thus, whether the distribution is uni- or bimodal is independent of $\alpha$ and $\gamma$ in this limit.\\
    \item There is always an extremum at $x=0$, which is either unique and a maximum (unimodality), or a minimum accompanied by two maxima symmetrically placed around it (bimodality) for $p$ is even. In the bimodal case, interestingly, the position of the maxima will not only depend on $\beta$ and $\kappa$ but also on the noise strength $\sigma^2$. In particular they will be at solutions to the following equation: $\tanh (2\beta x / \sigma^2) = \kappa x /\beta$, while $x\neq 0$.
\end{enumerate}

\subsubsection{Uni- and bimodality}
\label{app:uni_or_bimod_cond}
The distribution is unimodal, when the extremum at $x=0$ is a maximum, i.e. when $p''(0)\leq 0$, and bimodal in the complementary case.
The first derivative reads
\begin{equation}
    p'(x) = C \mathrm{e}^{-\frac{\kappa x^2}{\sigma^2}} \left[ \frac{2\beta}{\sigma^2} \sinh \left( \frac{2\beta}{\sigma^2} x \right) - \frac{2\kappa}{\sigma^2} x \cosh \left( \frac{2\beta}{\sigma^2} x \right) \right],
\end{equation}
(where $C$ is a positive constant factor) which clearly obeys $p'(0)=0$.
The second derivative is
\begin{equation}
    p''(x) = C' \mathrm{e}^{-\frac{\kappa x^2}{\sigma^2}} \left[  
    \frac{4\beta^2}{\sigma^4} \cosh \left( \frac{2\beta}{\sigma^2} x \right) -
     \frac{4\beta\kappa}{\sigma^4} x \sinh \left( \frac{2\beta}{\sigma^2} x \right) -
      \frac{2\kappa}{\sigma^2} \cosh \left( \frac{2\beta}{\sigma^2} x \right) +
       \frac{4\kappa^2}{\sigma^4} x^2 \cosh \left( \frac{2\beta}{\sigma^2} x \right)
    \right].
\end{equation}
Evaluated at $x=0$, this leads to
\begin{equation}
    p''(x) = C' \left( \frac{4\beta^2}{\sigma^4} - \frac{2 \kappa}{\sigma^2} \right),
\end{equation}
with constant $C'>0$, which implies
\begin{align}
    &\text{unimodality:} \quad \kappa \geq \frac{2\beta^2}{\sigma^2},\\
    &\text{bimodality:} \quad\,\,\,\, \kappa < \frac{2\beta^2}{\sigma^2}.
\end{align}

\section{Stationary Distribution for $\alpha\gg\kappa,\,\beta$ and $\gamma\sigma_N^2 \nrightarrow 0$}
\label{app:alpha_large}
Recall the following equation from Appendix~\ref{app:change_of_variables}, where the third and fourth line are added anew here:
\begin{align}
    \label{eq: change_var_xdot_withy2}
    \dot{x} &= -\kappa x + \beta \tanh (\gamma (y + \alpha x)) + \sigma_N \xi_t^N - \sigma_V \xi_t^V \nonumber\\
    &= -\kappa x + \beta \tanh (\gamma (\alpha \int_0^t \mathrm{e}^{-\alpha (t-s)} \dot{x}(s) \,\dd s +\alpha \sigma_V \int_{0}^t \mathrm{e}^{-\alpha (t-s)} \xi^V_s \,\dd s)) + \sigma_N \xi_t^N - \sigma_V \xi_t^V \\
    & \underset{\text{large}}{\overset{\alpha}{\approx}} -\kappa x + \beta \tanh (-\gamma\kappa x + \gamma\beta \tanh (\gamma (y + \alpha x)) +  \gamma \alpha\sigma_N \int_0^t \mathrm{e}^{-\alpha (t-s)} \xi_s^N \,\dd s) + \sigma_N \xi_t^N - \sigma_V \xi_t^V \\
    &\underset{\text{small}}{\overset{\beta, \, \kappa}{\approx}} -\kappa x + \beta \tanh (-\gamma(\kappa x - \beta \xi^\textup{tele}_t) +  \gamma \alpha\sigma_N \int_0^t \mathrm{e}^{-\alpha (t-s)} \xi_s^N \,\dd s) + \sigma_N \xi_t^N - \sigma_V \xi_t^V  + \mathcal{O} (\beta^3),
\end{align}
where in the third line it was used that $\alpha \, \mathrm{e}^{-\alpha (t-s)}\overset{\alpha\to\infty}{\longrightarrow} \delta (t-s) \,\,\forall t>s$, where $\delta (t-s)$ is a Dirac-delta .
The encapsulated hyperbolic tangent may be approximated by auto-correlated telegraphic noise, $\xi^\textup{tele} \in \{\pm 1\}$, as all other correction terms would be $\sim \mathcal{O} (\beta^3)$. Its auto-correlation structure and its effect are determined in the next subsection.
When $\kappa$, $\beta$ small, the hyperbolic tangent may be replaced by its (first order) Taylor expansion:
\begin{equation}
    \dot{x} \overset{\kappa\gamma x \ll 1}{\approx} -\kappa x - \beta\gamma(\kappa x - \beta \xi^\textup{tele}_t) \frac{1}{\cosh^2 (\gamma \alpha\sigma_N \int_0^t \mathrm{e}^{-\alpha (t-s)} \xi_s^N \,\dd s)} + \beta \tanh (\gamma \alpha\sigma_N \int_0^t \mathrm{e}^{-\alpha (t-s)} \xi_s^N \,\dd s) + \sigma_N \xi_t^N - \sigma_V \xi_t^V + \mathcal{O} (\beta^2).
\end{equation}
The term involving the hyperbolic cosine may be replaced by its average, owing to the fact that its argument is fast compared to $x$. The expectation of said term is calculated in Eq.~\eqref{eq: expectation_cosh^2} later in this section via the Furutsu-Novikov Theorem and reads ($X$ is the argument of the cosh in the previous equation)
\begin{equation}
    \bigg\langle \frac{1}{\cosh^2 (X)} \bigg\rangle = \frac{2}{\gamma\sigma_N} \frac{1}{\sqrt{\pi \alpha}} + \mathcal{O} \left(\frac{1}{\alpha^{3/2}}\right),
\end{equation}
such that the equation can be approximated as
\begin{align}
    \dot{x} 
    &\approx -\kappa x \left( 1 + \frac{2\beta}{\sigma_N \sqrt{\pi\alpha}} \right) +\frac{2\beta^2}{\sigma_N \sqrt{\pi \alpha}} \xi^\textup{tele} + \beta \tanh (\gamma \alpha\sigma_N \int_0^t \mathrm{e}^{-\alpha (t-s)} \xi_s^N \,\dd s) + \sigma_N \xi_t^N - \sigma_V \xi_t^V \\ \label{eq:D7}
    &= -\kappa_\text{eff\,} x + \frac{2\beta^2}{\sigma_N \sqrt{\pi \alpha}} \xi^\textup{tele} + \beta \tanh (\gamma \alpha\sigma_N \int_0^t \mathrm{e}^{-\alpha (t-s)} \xi_s^N \,\dd s) + \sigma_N \xi_t^N - \sigma_V \xi_t^V,\,\, \text{where}\,\, \kappa_\text{eff}=\kappa \left( 1 + \frac{2\Theta}{\sqrt{\pi}} \right),
\end{align}
where we have introduced the notation $\Theta:=\beta/\sigma_N \sqrt{\alpha}$.

When $\gamma\sigma_N^2 \nrightarrow 0$, while $\alpha\gg\kappa>0$, the hyperbolic tangent term becomes an auto-correlated/coloured telegraphic noise, $\xi^\textup{tele}_t \in \{\pm1\}$. The auto-correlation decay will be inherited in a non-trivial way from the OU process, which is derived in the next subsection.

The mispricing Langevin equation, Eq.\eqref{eq: change_var_xdot_withy2}, then reads
\begin{align}
    \dot{x} &= -\kappa_\text{eff\,} x + \beta \xi_t^\textup{tele} + \frac{2\beta^2}{\sigma_N \sqrt{\pi \alpha}} \xi^\textup{tele} + \sigma_N \xi_t^N - \sigma_V \xi_t^V \\
    &= -\kappa_\text{eff\,} x + \beta_\text{eff\,} \xi_t^\textup{tele} + \sigma_N \xi_t^N - \sigma_V \xi_t^V,\,\, \text{where} \,\, \beta_\text{eff\,} = \beta \left( 1 +  \frac{2\Theta}{\sqrt{\pi}}  \right).
\end{align}

This system has two potential steady states, $x=\pm \frac{\beta_\text{eff\,}}{\kappa_\text{eff\,}}$, when disregarding white noise. So at $x=\pm\frac{\beta_\text{eff\,}}{\kappa_\text{eff\,}}$ the process (disregarding white noise) is stable when the telegraphic noise is $\pm\beta_\text{eff\,}$ but it can get de-stabilised or perturbed by telegraphic noise of the opposite sign. Thus, the switching behaviour between the two possible states of the telegraphic process needs to be studied.

\subsection{Autocovariance Telegraphic Noise}

In this section the autocovariance of the telegraphic noise will be derived.

First, let $Y_t$ be the underlying OU-process. Henceforth, it is assumed that $Y$ is stationary, such that
\begin{equation}
    \langle Y_t Y_s \rangle \sim \mathrm{e}^{-\alpha |t-s|}.
\end{equation}
Further, for the OU-process is a Gaussian process, $Y\sim \mathcal{N} (0, \frac{\sigma^2}{2\alpha})$, where in this section $\sigma = \gamma\alpha\sigma_N$ is the constant prefactor of the white noise within the hyperbolic tangent.
Further note that in this limit
\begin{equation}
    \xi_t^\textup{tele}=1 \quad \Leftrightarrow\quad Y_t>0.
\end{equation}
Assuming $t>0$ and $\xi_s^\textup{tele}=1$, it follows that
\begin{align}
    \langle \xi_t^\textup{tele} \xi_s^\textup{tele} \rangle &= \mathbb{P} [\xi_t^\textup{tele}=1 | \xi_s^\textup{tele}=1] - \mathbb{P} [\xi_t^\textup{tele}=1 | \xi_s^\textup{tele}=-1] \\
    &= \mathbb{P} [Y_t>0 | Y_s>0] - \mathbb{P} [Y_t>0 | Y_s<0].
\end{align}

Using the result from the bivariate centered Gaussian distribution with correlation $\rho$ and variances $\sigma_1$ and $\sigma_2$, that
\begin{align}
    &\mathbb{P} [Y_t>0 ,\, Y_s>0] = \frac{1}{4} + \frac{1}{2\pi} \arcsin (\rho) \\
    &\mathbb{P} [Y_t>0 ,\, Y_s<0] = \frac{1}{2\pi} \arccos (\rho) \\
    &\mathbb{P} [Y_s>0] = \mathbb{P} [Y_s<0] = \frac{1}{2},
\end{align}
and with $\arcsin (x) - \arccos (x) = 2\arcsin (x)-\frac{\pi}{2}$, it follows that
\begin{equation}
    \langle \xi_t^\textup{tele} \xi_s^\textup{tele} \rangle = 2 \left(\frac{1}{4} + \frac{1}{2\pi}(2\arcsin (\mathrm{e}^{-\alpha |t-s|}) -\frac{\pi}{2})\right) = \frac{2}{\pi} \arcsin (\mathrm{e}^{-\alpha |t-s|}).
\end{equation}
Thus, the switching rate between the two steady states scales $\sim \alpha / \pi$ for large $\alpha$. When $\kappa\gg\lambda$, where $\lambda$ is the switching rate of the telegraphic noise, i.e. when the relaxation to the steady state is faster than the switching between the two modes, the dynamics
has enough time to relax to the steady states, resulting in two clearly distinguishable modes in distribution. In the opposite
case, when switching is in the same order of magnitude or even faster, the two modes of $p(x)$ become indistinguishable and are washed out by the noise. In the case where $\alpha\gg\kappa \sim\lambda$, the distribution is thus unimodal and centered around $x=0$ and Gaussian. Thus, we only need to calculate the second moment $\langle x^2 \rangle$ to determine $p(x)$.

\subsection{Variance of the Process}
The integrated version of the Langevin equation (using an integrating factor $\mathrm{e}^{\kappa_\text{eff\,} t}$) reads
\begin{equation}
    x(t) = \int_0^t \mathrm{e}^{-\kappa_\text{eff\,} (t-s)} \left[ \beta_\text{eff\,} \xi_s^\textup{tele} + \sigma_N \xi_s^N - \sigma_V \xi_s^V \right] \, \dd s =: A(t) + B(t) + C(t),
\end{equation}
where the terms $A$, $B$, $C$ are defined by the three integrals.
The variance of the process is thus given by
\begin{equation}
    \langle x^2 \rangle = \langle A^2 \rangle + \langle B^2 \rangle + \langle C^2 \rangle + 2 \langle AB \rangle = \langle A^2 \rangle + \frac{\sigma_N^2 + \sigma_V^2}{2\kappa_\text{eff\,}} + 2 \langle AB \rangle
\end{equation}
because all terms are centered.

Further, we know that

\begin{align}
    \langle A^2 \rangle &= \langle \int_0^t \beta_\text{eff\,} \mathrm{e}^{-\kappa_\text{eff\,} (t-s)} \xi_s^\textup{tele} \,\dd s \int_0^t \beta_\text{eff\,} \mathrm{e}^{-\kappa_\text{eff\,} (t-s')} \xi_{s'}^\textup{tele} \,\dd s' \rangle \\
   &= \int_0^t \int_0^t \beta^2_\text{eff\,} \mathrm{e}^{-\kappa_\text{eff\,} (2t-s-s')} \langle \xi_s^\textup{tele} \xi_{s'}^\textup{tele} \rangle \,\dd s\dd s' \\
   &= \int_0^t \int_0^t \beta^2_\text{eff\,} \mathrm{e}^{-\kappa_\text{eff\,} (2t-s-s')} \frac{2}{\pi} \arcsin (\mathrm{e}^{-\alpha |s'-s|}) \,\dd s\dd s'.
\end{align}
Defining $u=t-s$ and $u'=t-s'$, such that $\dd s\dd s' = \dd u \dd u'$, the integral can be rewritten as
\begin{equation}
     \langle A^2 \rangle = \frac{2\beta^2_\text{eff\,}}{\pi} \int_0^t \int_0^t \mathrm{e}^{-\kappa_\text{eff\,} (u+u')} \arcsin (\mathrm{e}^{-\alpha |u'-u|}) \, \dd u \,\dd u',
\end{equation}
which becomes the following in the long-time limit
\begin{equation}
     \langle A^2 \rangle = \frac{2\beta^2_\text{eff\,}}{\pi} \int_0^\infty \int_0^\infty \mathrm{e}^{-\kappa_\text{eff\,} (u+u')} \arcsin (\mathrm{e}^{-\alpha |u'-u|}) \, \dd u \,\dd u'.
\end{equation}
With new variables $x=1/2 (u+u')$ and $y=1/2 (u-u')$, such that $u=x-y$, $u'=x+y$ and $\dd u\,\dd u' = 2 \dd x\,\dd y$, the integral reads
\begin{align}
    \langle A^2 \rangle &= \frac{4\beta^2_\text{eff\,}}{\pi} \int_0^\infty \int_{-x}^x \mathrm{e}^{-2\kappa_\text{eff\,} x} \arcsin (\mathrm{e}^{-2\alpha |y|}) \,\dd y\,\dd x\\
    &= \frac{8\beta^2_\text{eff\,}}{\pi} \int_0^\infty \mathrm{e}^{-2\kappa_\text{eff\,} x} \left( \int_{0}^x \arcsin (\mathrm{e}^{-2\alpha y}) \,\dd y \right) \, \dd x,
\end{align}
where the last step holds because the arcsine is even in $y$. The integral bounds have changed because $u, u'\in [0, \,\infty)$, so $u=x-y\geq 0$ and $u'=x+y\geq 0$ requires $|y|\leq x$, such that finally $x\in [0,\,\infty)$ and $y\in [-x, x]$.
Changing the order of integration, one finds
\begin{align}
    \langle A^2 \rangle &= \frac{8\beta^2_\text{eff\,}}{\pi} \int_0^\infty \arcsin (\mathrm{e}^{-2\alpha y}) \left( \int_y^\infty \mathrm{e}^{-2\kappa_\text{eff\,} x} \,\dd x \right) \,\dd y \\
    &= \frac{4\beta^2_\text{eff\,}}{\pi\kappa_\text{eff\,}} \int_0^\infty \arcsin (\mathrm{e}^{-2\alpha y}) \mathrm{e}^{-2\kappa_\text{eff\,} y} \,\dd y\\
    &= \frac{4\beta^2_\text{eff\,}}{\pi\kappa_\text{eff\,}} \frac{\sqrt{\pi}}{4\kappa_\text{eff\,}}\left( \sqrt{\pi} - \frac{\Gamma (\frac{\alpha+\kappa_\text{eff\,}}{2\alpha})}{\Gamma (\frac{2\alpha +\kappa_\text{eff\,}}{2\alpha})} \right) =
    \frac{\beta^2_\text{eff\,}}{\sqrt{\pi}\kappa_\text{eff\,}^2} \left( \sqrt{\pi} - \frac{\Gamma (\frac{\alpha+\kappa_\text{eff\,}}{2\alpha})}{\Gamma (\frac{2\alpha +\kappa_\text{eff\,}}{2\alpha})} \right) .
\end{align}

The last outstanding term is
\begin{equation}
    \langle A B \rangle = \big\langle \int_0^t \mathrm{e}^{-\kappa_\text{eff\,} (t-s)} \beta_\text{eff\,} \xi_s^\textup{tele} \,\dd s \int_0^t \mathrm{e}^{-\kappa_\text{eff\,} (t-s')} \sigma_N \xi_{s'}^N \,\dd s' \big\rangle.
\end{equation}
In order to calculate the covariance, we write the telegraphic noise again in its non-simplified form using the hyperbolic tangent:
\begin{equation}
    \langle A B \rangle = \int_0^t \int_0^t \mathrm{e}^{-\kappa_\text{eff\,} (2t-s-s')} \beta_\text{eff\,} \sigma_N \langle \tanh (\gamma\alpha\sigma_N \int_0^s \mathrm{e}^{-\alpha (s-r)} \xi_r^N \,\dd r) \xi_{s'}^N \rangle \,\dd s'\,\dd s
\end{equation}
The expectation is over a functional  of the white noise $\xi_t^N$ multiplied by the same noise. For such expectations, the Novikov theorem holds, which states for a functional $\mathcal{F}[\eta]$ of some Gaussian noise $\eta_t$ that
\begin{equation}
    \langle \mathcal{F} [\eta] \eta_t  \rangle = \int \langle \frac{\delta \mathcal{F}}{\delta \eta_s} \rangle \langle \eta_s \eta_t \rangle \,\dd s,
\end{equation}
where $\delta$ in this expression refers to the functional derivative \cite{ishimaru1978wave}.
In the case of Gaussian white noise, i.e. $\langle \eta_s \eta_t \rangle = \delta (s-t)$ (in this term $\delta$ is the Dirac-delta), the theorem simplifies to
\begin{equation}
    \langle \mathcal{F} [\eta] \eta_t  \rangle =  \bigg\langle \frac{\delta \mathcal{F}}{\delta \eta_t} \bigg\rangle .
\end{equation}
Let us define
\begin{equation}
    X_s [\xi^N] = \gamma \alpha \sigma_N \int_0^s \mathrm{e}^{-\alpha (s-r)} \xi_r^N \,\dd r,
\end{equation}
such that $\mathcal{F} [\xi^N] = \tanh(X_s[\xi^N])$. The functional derivative then is
\begin{equation}
    \frac{\delta \mathcal{F}[\xi]}{\delta \xi_{s'}^N} = \frac{\dd \tanh (X_s)}{\dd X_s} \frac{\delta X_s}{\delta \xi_{s'}^N} = (1-\tanh^2 (X_s)) \frac{\delta X_s}{\delta \xi_{s'}^N} = \frac{1}{\cosh^2 (X_s)} \alpha\gamma\sigma_N \mathrm{e}^{-\alpha (s-s')} \Theta (s-s').
\end{equation}
Note that
\begin{equation}
    \bigg\langle \frac{\delta \mathcal{F}[\xi]}{\delta \xi_{s'}^N} \bigg\rangle \sim \bigg\langle \frac{1}{\cosh^2 (X_s)} \bigg\rangle
\end{equation}
and that $X$ is a Gaussian process, which converges to $\mathcal{N}(0, w)$ in distribution, where in the stationary limit $w=\frac{\gamma^2\alpha\sigma_N^2}{2}$ is the stationary variance.
Therefore,
\begin{equation}
    \bigg\langle \frac{1}{\cosh^2 (X)} \bigg\rangle = \frac{1}{\sqrt{2\pi w}} \int_{-\infty}^\infty \mathrm{e}^{\frac{-X^2}{2w}} \frac{1}{\cosh^2 (X)} \,\dd X.
\end{equation}
In the large-$w$ limit, which is the case considered in this section, the integral is dominated by $X=\mathcal{O}(1)$ because of $\cosh^{-2} (X)$. Consequently $X^2/(2w)\ll 1$, such that the exponential term may be approximated by its Taylor series:
\begin{equation}
    \bigg\langle \frac{1}{\cosh^2 (X_s)} \bigg\rangle \approx \frac{1}{\sqrt{2\pi w}} \int_{-\infty}^\infty \left( 1 - \frac{X^2}{2w} + \dots \right) \frac{1}{\cosh^2 (X_s)} \,\dd X.
\end{equation}
Keeping only the leading order, this results in
\begin{equation}
    \label{eq: expectation_cosh^2}
    \bigg\langle \frac{1}{\cosh^2 (X_s)} \bigg\rangle = \frac{1}{\sqrt{2\pi w}} \int_{-\infty}^\infty \frac{1}{\cosh^2 (X)} \,\dd X + \mathcal{O} \left(\frac{1}{w^{3/2}}\right) = \sqrt{\frac{2}{\pi w}} + \mathcal{O} \left(\frac{1}{w^{3/2}}\right) = \frac{2}{\gamma\sigma_N} \frac{1}{\sqrt{\pi \alpha}} + \mathcal{O} \left(\frac{1}{w^{3/2}}\right).
\end{equation}
Consequently, the Furutsu-Novikov theorem yields approximately 
\begin{equation}
    \bigg\langle \frac{\delta \mathcal{F}[\xi]}{\delta \xi_{s'}^N} \bigg\rangle \approx \frac{2}{\gamma\sigma_N} \frac{1}{\sqrt{\pi \alpha}} \alpha\gamma\sigma_N \mathrm{e}^{-\alpha (s-s')} \Theta (s-s') = 2 \sqrt{\frac{\alpha}{\pi}} \mathrm{e}^{-\alpha (s-s')} \Theta (s-s'),
\end{equation}
which renders the cross-correlation to be
\begin{equation} \langle AB \rangle \approx
    2\beta_\text{eff\,}\sigma_N \sqrt{\frac{\alpha}{\pi}} \int_0^t \int_0^t \mathrm{e}^{-\kappa_\text{eff\,} (2t-s-s')} \mathrm{e}^{-\alpha (s-s')} \Theta (s-s') \, \dd s \, \dd s'.
\end{equation}
Using the same substitution for $u$, $u'$ as for the term $\langle A^2 \rangle$ and taking again the stationary limit $t\to \infty$, leads to
\begin{align}
    \langle AB \rangle &\approx 2\beta_\text{eff\,}\sigma_N \sqrt{\frac{\alpha}{\pi}} \int_0^\infty \int_0^\infty \mathrm{e}^{-\kappa_\text{eff\,} (u+u')} \mathrm{e}^{-\alpha (u-u')} \Theta (u-u') \, \dd u' \, \dd u \\
    &=2\beta_\text{eff\,}\sigma_N \sqrt{\frac{\alpha}{\pi}} \int_0^\infty \int_0^u \mathrm{e}^{-u(\kappa_\text{eff\,}+\alpha)} \mathrm{e}^{u'(\alpha-\kappa_\text{eff\,})} \, \dd u' \, \dd u \\
    &=2\beta_\text{eff\,}\sigma_N \sqrt{\frac{\alpha}{\pi}} \frac{1}{\alpha-\kappa_\text{eff\,}}\int_0^\infty  \mathrm{e}^{-u(\kappa_\text{eff\,}+\alpha)} \left(\mathrm{e}^{u(\alpha-\kappa_\text{eff\,})}-1\right) \, \dd u \\
    &=2\beta_\text{eff\,}\sigma_N \sqrt{\frac{\alpha}{\pi}} \frac{1}{\alpha-\kappa_\text{eff\,}}\int_0^\infty  \left(\mathrm{e}^{-2u\kappa_\text{eff\,}} - \mathrm{e}^{-u(\alpha+\kappa_\text{eff\,})}\right) \, \dd u
    \\
    &=2\beta_\text{eff\,}\sigma_N \sqrt{\frac{\alpha}{\pi}} \frac{1}{\alpha-\kappa_\text{eff\,}} \left(\frac{1}{2\kappa_\text{eff\,}} - \frac{1}{\alpha+\kappa_\text{eff\,}}\right) \\
    &=\beta_\text{eff\,}\sigma_N \sqrt{\frac{\alpha}{\pi}} \frac{1}{\kappa_\text{eff\,}(\alpha+\kappa_\text{eff\,})}.
\end{align}

Knowing this, the variance of the process $x$ is
\begin{equation}
    \langle x^2 \rangle = \langle A^2 \rangle + \langle B^2 \rangle + \langle C^2 \rangle + 2\langle AB \rangle.
\end{equation}

This concludes our derivation of the stationary distribution in the case where $\alpha\gamma^2\sigma_N^2$ is large and $\alpha\gg\kappa$, which reads
\begin{align}
    p(x) &= \mathcal{N} (0,\, \langle x^2 \rangle) = \mathcal{N} \left(0,\, \frac{\sigma_N^2 + \sigma_V^2}{2\kappa_\text{eff\,}} + \frac{\beta_\text{eff\,}^2}{\sqrt{\pi}\kappa_\text{eff\,}^2} \left( \sqrt{\pi} - \frac{\Gamma (\frac{\alpha+\kappa_\text{eff\,}}{2\alpha})}{\Gamma (\frac{2\alpha+\kappa_\text{eff\,}}{2\alpha})} \right) + 2\beta_\text{eff\,}\sigma_N \sqrt{\frac{\alpha}{\pi}} \frac{1}{\kappa_\text{eff\,}(\alpha+\kappa_\text{eff\,})} \right) \\
    &= \mathcal{N} \left(0,\,  \frac{\sigma_N^2 + \sigma_V^2}{2\kappa_\text{eff\,}} + \frac{\beta^2}{\sqrt{\pi}\kappa^2} \left( \sqrt{\pi} - \frac{\Gamma (\frac{\alpha+\kappa_\text{eff\,}}{2\alpha})}{\Gamma (\frac{2\alpha+\kappa_\text{eff\,}}{2\alpha})} \right) + 2\beta\sigma_N \sqrt{\frac{\alpha}{\pi}} \frac{1}{\kappa (\alpha + \kappa_\text{eff\,})} \right),
\end{align}
where $\mathcal{N}$ refers to the Gaussian distribution and, again,
\begin{equation}
    \kappa_\text{eff\,} = \kappa \left( 1 + \frac{2\beta}{\sigma_N \sqrt{\pi\alpha}} \right)\quad\text{and}\quad \beta_\text{eff\,} = \beta \left( 1 + \frac{2\beta}{\sigma_N \sqrt{\pi\alpha}}\right) .
\end{equation}
Keeping only terms up to quadratic order in $\beta$, this finally leads to
\begin{align}
    p(x) &= \mathcal{N} \left(0,\,  \frac{\sigma_N^2 + \sigma_V^2}{2\kappa_\text{eff\,}} + \frac{\beta^2}{\sqrt{\pi}\kappa^2} \left( \sqrt{\pi} - \sqrt{\pi} (1- \frac{\ln (2) \kappa}{\alpha}) \right) + 2\beta\sigma_N \sqrt{\frac{\alpha}{\pi}} \frac{1}{\kappa (\alpha + \kappa_\text{eff\,})} + \mathcal{O} (\beta^3) \right) \nonumber\\
    &= \mathcal{N} \left(0,\,  \frac{\sigma_N^2 + \sigma_V^2}{2\kappa_\text{eff\,}} + \frac{\ln (2) \beta^2}{\kappa \alpha} + 2\beta\sigma_N \sqrt{\frac{\alpha}{\pi}} \frac{1}{\kappa (\alpha + \kappa_\text{eff\,})} + \mathcal{O} (\beta^3) \right),
    \label{eq:misp_dist_wNovikov_final}
\end{align}
which is justified by the Taylor expansion involving the fractions of Gamma-functions in the following subsection.

\subsection{Leading orders of $\langle A^2 \rangle$}
We have found (in the limit $\alpha \gg \kappa$) the closed-form solution
\begin{equation}
    \langle A^2 \rangle = \frac{\beta^2}{\sqrt{\pi}\kappa^2} \left( \sqrt{\pi} - \frac{\Gamma (\frac{\alpha+\kappa_\text{eff}}{2\alpha})}{\Gamma (\frac{2\alpha+\kappa_\text{eff}}{2\alpha})} \right).
\end{equation}
Let us determine the leading order in $\Gamma$ when $\alpha \gg \kappa$.

Defining $\epsilon = \frac{\kappa}{2 \alpha}$, the argument of the Gamma-fuction in the numerator is $\frac{1}{2} + \epsilon$ and of the denominator $1+\epsilon$. $\epsilon\ll 1$ when $\alpha \gg \kappa$, such that the expansions of the Gamma-function near 1 and $1/2$ may be used:
\begin{align}
    \Gamma (1+\epsilon) &= 1 - \Tilde{\gamma} \epsilon + \mathcal{O} (\epsilon^2),\\
    \Gamma (1/2 + \epsilon) &= \underbrace{\Gamma (\frac{1}{2})}_{=\sqrt{\pi}} \left( 1 + \Psi (\frac{1}{2}) \epsilon + \mathcal{O} (\epsilon^2) \right) = \sqrt{\pi} \left( 1-(\Tilde{\gamma} + 2\ln (2))\epsilon + \mathcal{O} (\epsilon^2) \right),
\end{align}
where $\Tilde{\gamma}$ is the Euler-Mascheroni constant and $\Psi$ the digamma function. $\Psi$ is connected to $\Gamma$ via its derivative: $\frac{\dd}{\dd x}\Gamma (x) = \Gamma (x) \Psi (x)$. It is $\Psi (1) = -\Tilde{\gamma}$ and $\Psi (1/2) = -\Tilde{\gamma} - 2 \ln (2)$.

Taking the ratio
\begin{equation}
    \frac{\Gamma (1/2 + \epsilon)}{\Gamma (1+\epsilon)} \approx \frac{\sqrt{\pi} (1-(\Tilde{\gamma} + 2\ln (2))\epsilon)}{1-\Tilde{\gamma}\epsilon}
\end{equation}
and using the first-order approximation
\begin{equation}
    \frac{1-a \epsilon}{1-b \epsilon} \approx 1+ (b-a)\epsilon,
\end{equation}
leads to
\begin{equation}
    \frac{\Gamma (1/2 + \epsilon)}{\Gamma (1+\epsilon)} = \sqrt{\pi} \left( 1 - 2\ln (2)\epsilon \right) + \mathcal{O} (\epsilon^2) = \sqrt{\pi} \left( 1 - \frac{\ln (2)\kappa_\text{eff}}{\alpha} \right) + \mathcal{O} (\epsilon^2).
\end{equation}

\section{Stationary Distribution for $\alpha,\,\beta \gg \kappa$ and $\gamma\sigma_N^2 \nrightarrow 0$}
\label{app:alpha_beta_large}
In this section, we motivate and derive another result  in the limit, where both $\alpha\gg\kappa$ (as in the previous section) but also $\beta\gg \kappa$ (while $\gamma \sigma_N^2 \nrightarrow 0$ for the calculation to hold). In this limit, because $\beta\gg\kappa$, the trend signal distribution $p(M)$ is bimodal. However, unlike claims in the literature this does \textit{not} automatically imply a bimodality of the mispricing distribution $p(\delta)$. $p(M)$ in this limit is derived here and a motivation for the P-bifurcation of the mispricing distribution $p(\delta)$ that is bimodal only 'later', i.e. for even larger values of $\beta$, is given.

Recall the change of variables stated in Appendix~\ref{app:change_of_variables}, $M=y+\alpha x$, from which it follows using eq.~\eqref{eq:change_var_xdot} and eq.~\eqref{eq:change_var_ydot} that
\begin{align}
    \dot{x} & = -\kappa x + \beta \tanh (\gamma M) + \sigma_N\xi_t^N - \sigma_V\xi_t^V \nonumber\\
    \dot{M} &= \dot{y} + \alpha \dot{x} \nonumber\\
    & = -\alpha M + \alpha \sigma_V\xi_t^V + \alpha (\kappa x + \beta \tanh (\gamma M) + \sigma_N\xi_t^N - \sigma_V\xi_t^V) \nonumber\\
    & = -\alpha M + \alpha\kappa x + \alpha\beta \tanh (\gamma M) + \alpha\sigma_N \xi_t^N
\end{align}

\subsubsection{Quasi-static Assumption for $x$}
Owing to the fact that $\alpha\gg\kappa$, $M$ relaxes on time scales much faster than $x$, such that a quasi-static assumption for $x$ is justified relative to $M$ (we will show in the next section, why and when this assumption breaks down when $\beta$ is increased). The following conditional Fokker-Planck equation can be written down:
\begin{equation}
    \frac{\partial p(M|x)}{\partial t} = -\frac{\partial}{\partial M} \left[ (-\alpha M + \alpha\kappa x + \alpha\beta \tanh (\gamma M)) p(M|x) \right] + \frac{\sigma_N^2 \alpha^2}{2} \frac{\partial^2 p(M|x)}{\partial M^2}.
    \label{eq:cond_FPE_M_given_X}
\end{equation}
The stationary solution ($\frac{\partial p(M|x)}{\partial t}=0$) is given through the Maxwell-Boltzmann {\it ansatz} and reads
\begin{align}
    p(M|x) & = \frac{1}{A(x)} \exp \left( \frac{2}{\sigma_N^2 \alpha^2} (-\frac{\alpha}{2} M^2 + \alpha\kappa M x + \frac{\alpha\beta}{\gamma} \ln (\cosh (\gamma M))) \right) \\ \label{eq:E4}
    &= \frac{1}{A(x)} \cosh (\gamma M)^\frac{2\beta}{\sigma_N^2\alpha\gamma} \,\mathrm{e}^{-\frac{1}{\sigma_N^2 \alpha}M^2 + \frac{2\kappa}{\sigma_N^2 \alpha} M x},
\end{align}
where $A(x)$ is the normalisation. Note that for $x=0$, $p(M|x)$ becomes bimodal when $\beta \gamma > 1$.

Assuming that this stationary distribution is reached very quickly, such that $x$ hardly moves, one can make progress and compute how the dynamics of $x$ itself is affected by the trend $M$. Within this quasi-static assumption, we can  replace the hyperbolic tangent term by its expectation in the Langevin evolution of $x$:
\begin{equation}
    \dot{x}  \approx -\kappa x + \beta ( \mathbb{E}[ \tanh (\gamma M)|x] + \tilde{\xi}_t^{\text{tele}}) + \sigma_N\xi_t^N - \sigma_V\xi_t^V,
\end{equation}
where the telegraphic noise $\xi_t^{\text{tele}}$ has been separated into its mean
\begin{align}
    \mathbb{E} [\tanh (\gamma M)|x] &= \int_{-\infty}^{\infty} p(M|x) \tanh (\gamma M) \,\mathrm{d} M \\
    &= \frac{1}{A(x)} \int_{-\infty}^{\infty} \cosh (\gamma M)^\frac{2\beta}{\sigma_N^2\alpha\gamma} \exp \left( -\frac{1}{\sigma_N^2 \alpha} M^2 + \frac{2\kappa}{\sigma_N^2 \alpha} M x \right) \tanh (\gamma M) \,\mathrm{d}M.
\end{align}
and a mean-zero contribution $\tilde{\xi}_t^{\text{tele}}$ with variance $\beta^2$. 

The normalisation $A(x)$ reads
\begin{equation}
    A(x) = \int_{-\infty}^\infty  \cosh (\gamma M)^\frac{2\beta}{\sigma_N^2\alpha\gamma} \,\mathrm{e}^{-\frac{1}{\sigma_N^2 \alpha}M^2 + \frac{2\kappa}{\sigma_N^2 \alpha} M x} \,\mathrm{d}M.
\end{equation}
Now, after making the change of variable $M = \sigma_N \sqrt{\alpha} u$ one can see that when $\gamma \sigma_N \sqrt{\alpha} \gg 1$, which we will assume henceforth, 
one can replace in the integral $\cosh(\gamma M)$ by $\exp(\gamma |M|)/2$, up to a correction of order $(\gamma \sigma_N \sqrt{\alpha})^{-1}$ in the final result:
\begin{align} \label{eq:E8}
    A (x) &\approx \left(\frac{1}{2}\right)^\frac{2\beta}{\sigma_N^2\alpha\gamma} \int_{-\infty}^\infty  \mathrm{e}^{\frac{2\beta}{\sigma_N^2\alpha} |M|} \,\mathrm{e}^{-\frac{1}{\sigma_N^2 \alpha}M^2 + \frac{2\kappa}{\sigma_N^2 \alpha} M x} \,\mathrm{d}M \nonumber\\
    &= \left(\frac{1}{2}\right)^{\frac{2\beta}{\sigma_N^2\alpha\gamma}+1} \sqrt{\alpha\sigma_N^2 \pi} \left[ \mathrm{e}^{\frac{(\beta+x\kappa)^2}{\alpha\sigma_N^2}} \left( 1 + \erf (\frac{\beta + x\kappa}{\sqrt{\alpha}\sigma_N}) \right) +
    \mathrm{e}^{\frac{(\beta-x\kappa)^2}{\alpha\sigma_N^2}} \left( 1 + \erf (\frac{\beta - x\kappa}{\sqrt{\alpha}\sigma_N}) \right)
    \right]
\end{align}

Using the same approximations, the expectation $\mathbb{E} [\tanh (\gamma M)]$ is
\begin{align}
    \mathbb{E} [\tanh (\gamma M)|x] &\approx \frac{1}{A (x)} \left(\frac{1}{2}\right)^\frac{2\beta}{\sigma_N^2\alpha\gamma} \int_{-\infty}^\infty  \mathrm{e}^{\frac{2\beta}{\sigma_N^2\alpha} |M|} \,\mathrm{e}^{-\frac{1}{\sigma_N^2 \alpha}M^2 + \frac{2\kappa}{\sigma_N^2 \alpha} M x} \mathrm{sgn} (M)\,\mathrm{d}M \nonumber\\
    &=
    \frac{1}{A(x)} \left(\frac{1}{2}\right)^{\frac{2\beta}{\sigma_N^2\alpha\gamma}+1}  \sqrt{\alpha\sigma_N^2 \pi} \left[ \mathrm{e}^{\frac{(\beta+x\kappa)^2}{\alpha\sigma_N^2}} \left( 1 + \erf (\frac{\beta + x\kappa}{\sqrt{\alpha}\sigma_N}) \right) -
    \mathrm{e}^{\frac{(\beta-x\kappa)^2}{\alpha\sigma_N^2}} \left( 1 + \erf (\frac{\beta - x\kappa}{\sqrt{\alpha}\sigma_N}) \right) \right]
    \label{eq:norm_cond_density_Ax}
\end{align}
Plugging in the normalisation, we find
\begin{equation}
    \mathbb{E} [\tanh (\gamma M)|x] \approx
    \frac{\mathrm{e}^{\frac{2x\beta\kappa}{\alpha\sigma_N^2}}\left( 1 + \erf (\frac{\beta + x\kappa}{\sqrt{\alpha}\sigma_N}) \right) - \mathrm{e}^{-\frac{2x\beta\kappa}{\alpha\sigma_N^2}} \left( 1 + \erf (\frac{\beta - x\kappa}{\sqrt{\alpha}\sigma_N}) \right)}{\mathrm{e}^{\frac{2x\beta\kappa}{\alpha\sigma_N^2}} \left( 1 + \erf (\frac{\beta + x\kappa}{\sqrt{\alpha}\sigma_N}) \right) + \mathrm{e}^{-\frac{2x\beta\kappa}{\alpha\sigma_N^2}} \left( 1 + \erf (\frac{\beta - x\kappa}{\sqrt{\alpha}\sigma_N}) \right)}
\end{equation}
For $\kappa$ is small, the expectation is approximated by its (first-order) Taylor-expansion. Therefore, let
\begin{equation}
    E(z) = 1 + \erf (z) \quad \Rightarrow \quad E'(z) = \frac{2}{\sqrt{\pi}} \mathrm{e}^{-z^2}
\end{equation}
and expand
\begin{align}
    \mathrm{e}^{\pm \frac{2x\beta\kappa}{\alpha\sigma_N^2}} &= 1 \pm \frac{2x\beta}{\alpha\sigma_N^2} \kappa + \mathcal{O} (\kappa^2) \\
    E (\frac{\beta \pm x\kappa}{\sqrt{\alpha}\sigma_N}) &= E (\frac{\beta}{\sqrt{\alpha}\sigma_N}) \pm \frac{x}{\sqrt{\alpha}\sigma_N} \kappa E' (\frac{\beta}{\sqrt{\alpha}\sigma_N}) + \mathcal{O} (\kappa^2).
\end{align}
such that the products of the two terms scale as
\begin{align}
    A_1 &:= \mathrm{e}^{\frac{2x\beta\kappa}{\alpha\sigma_N^2}} E (\frac{\beta + x\kappa}{\sqrt{\alpha}\sigma_N}) = E (\frac{\beta}{\sqrt{\alpha}\sigma_N}) + \kappa \left[ \frac{2x\beta}{\alpha\sigma_N^2} E (\frac{\beta}{\sqrt{\alpha}\sigma_N}) + \frac{x}{\sqrt{\alpha}\sigma_N} E' (\frac{\beta}{\sqrt{\alpha}\sigma_N})\right] + \mathcal{O} (\kappa^2)
    \label{eq:Taylor_expansion_A1} \\
    A_2 &:= \mathrm{e}^{- \frac{2x\beta\kappa}{\alpha\sigma_N^2}} E (\frac{\beta - x\kappa}{\sqrt{\alpha}\sigma_N}) = E (\frac{\beta}{\sqrt{\alpha}\sigma_N}) - \kappa \left[ \frac{2x\beta}{\alpha\sigma_N^2} E (\frac{\beta}{\sqrt{\alpha}\sigma_N}) + \frac{x}{\sqrt{\alpha}\sigma_N} E' (\frac{\beta}{\sqrt{\alpha}\sigma_N})\right] + \mathcal{O} (\kappa^2).
    \label{eq:Taylor_expansion_A2}
\end{align}
One then finds that
\begin{equation}
    \frac{A_1-A_2}{A_1+A_2} = \frac{2\kappa \left[ \frac{2x\beta}{\alpha\sigma_N^2} E (\frac{\beta}{\sqrt{\alpha}\sigma_N}) + \frac{x}{\sqrt{\alpha}\sigma_N} E' (\frac{\beta}{\sqrt{\alpha}\sigma_N}) \right]}{2 E (\frac{\beta}{\sqrt{\alpha}\sigma_N})} + \mathcal{O} (\kappa^3) = \kappa \left( \frac{2x\beta}{\alpha\sigma_N^2} + \frac{x}{\sqrt{\alpha}\sigma_N} \frac{E'(\frac{\beta}{\sqrt{\alpha}\sigma_N})}{E(\frac{\beta}{\sqrt{\alpha}\sigma_N})} \right) + \mathcal{O} (\kappa^3),
\end{equation}
where the correction term is of order $\kappa^3$ by symmetry. This implies that, with $\Theta:=\beta/\sqrt{\alpha \sigma_N^2}$,
\begin{equation}
    \mathbb{E} [\tanh (\gamma M)|x] = \frac{2\kappa}{\beta} x \left[ \Theta^2 + \frac{1}{\sqrt{\pi}} \Theta  \frac{\mathrm{e}^{-\Theta^2}}{1 + \erf (\Theta)}  \right] + \mathcal{O} (\kappa^3).
\end{equation}

This finally means that an effective Langevin equation  for a generalised OU-type process with a modified mean-reversion speed may be written down, which reads
\begin{align}
    \dot{x} &= -\kappa x + \beta  ( \mathbb{E}[ \tanh (\gamma M)|x] + \tilde{\xi}_t^{\text{tele}}) + \sigma_N\xi_t^N - \sigma_V\xi_t^V \\
    &= -\kappa_\text{eff} x + \beta \tilde{\xi}_t^{\text{tele}} + \sigma_N\xi_t^N - \sigma_V\xi_t^V + \mathcal{O} (\kappa^3),
\end{align}
where 
\begin{equation} \label{eq:D7bis}
    \kappa_\text{eff} := \kappa Z(\Theta), \qquad Z(\Theta):=\left( 1 - 2\Theta^2 + \frac{2}{\sqrt{\pi}} \Theta  \frac{\mathrm{e}^{-\Theta^2}}{1 + \erf (\Theta)}  \right).
\end{equation}
In other words, the basic mechanism here is that a non-zero value of $x$ polarizes the trend $M$ in one direction, which in turn feeds back on $x$ itself and amplifies its bias. Note that Eq. \eqref{eq:D7bis} coincide with Eq. \eqref{eq:D7} when $\Theta \ll 1$, as expected. Interestingly $\kappa_\text{eff}$ first increases when $\Theta$ is small before decreasing and changing sign for $\Theta=\Theta_c \approx 0.797999$, see discussion below.

In order to write down the corresponding Fokker-Planck equation, the effective variance of the noise acting on $x$ is useful. 
This can be computed as (for all three noise sources have mean zero)
\begin{align}
    \sigma_x^2 &:= 2 \kappa \, \mathbb{E} \left[ \int_0^t \int_0^t \mathrm{e}^{-\kappa (2t-s-s')} \left( \beta \tilde{\xi}^{\text{tele}}_s + \sigma_N \xi^N_s - \sigma_V \xi^V_s \right) \left( \beta\tilde{\xi}^{\text{tele}}_{s'} + \sigma_N \xi^N_{s'} - \sigma_V \xi^V_{s'} \right) \,\dd s\,\dd s' \right]. \nonumber\\
    &= 2 \kappa \int_0^t \int_0^t \mathrm{e}^{-\kappa (2t-s-s')} \left(  \beta^2 \underbrace{\mathbb{E} [\tilde{\xi}^\textup{tele}_s \tilde{\xi}^\textup{tele}_{s'}]}_{\approx \frac{2 \ln (2)}{\alpha} \delta (s-s')} + 2\beta\sigma_N  \underbrace{\mathbb{E} [\tilde{\xi}^\textup{tele}_s \xi^N_{s'}]}_{= 2 \sqrt{\frac{\alpha}{\pi}} \mathrm{e}^{-\alpha (s-s')} \Theta (s-s')} +  \sigma_N^2 \underbrace{\mathbb{E} [\xi^N_s \xi^N_{s'}]}_{=\delta (s-s')} +  \sigma_V^2 \underbrace{\mathbb{E} [\xi^V_s \xi^V_{s'}]}_{=\delta (s-s')} \right) \,\dd s\,\dd s' \nonumber\\
    &\approx  {\sigma_N^2} \left(1 + \Xi^2 +  \frac{4}{\sqrt{\pi}}  \Theta + {2 \ln (2) \Theta^2}   \right), \qquad (\alpha \gg \kappa),
\end{align}
where the last two summands are due to the results in Eq.~\eqref{eq:misp_dist_wNovikov_final} and the preceding steps, and $\Xi^2:= \sigma_V^2/\sigma_N^2$ is the inverse of the well-known excess volatility ratio, see \cite{kurth2025revisiting}.

Thus, the following Fokker-Planck equation may be written down:
\begin{align}
     \frac{\partial p(x)}{\partial t} &= -\frac{\partial}{\partial x} [(-\kappa x + \beta \mathbb{E}[ \tanh (\gamma M)|x])p(x)] + \frac{\sigma_x^2}{2} \frac{\partial^2}{\partial x^2} p(x) \nonumber\\
     &= -\frac{\partial}{\partial x} [-\kappa_\textup{eff} x \, p(x)] + \frac{\sigma_x^2}{2} \frac{\partial^2}{\partial x^2} p(x).
\end{align}
The solution is derived with a Maxwell-Boltzmann {\it ansatz} as before
\begin{align}
    p(x) 
    &= \frac{1}{B} \exp \left(-\frac{Z(\Theta)\kappa}{\sigma_x^2} x^2 \right), 
    \qquad  B := \sqrt{\frac{\pi \sigma_x^2}{Z(\Theta)\kappa}}.
\end{align}
Assuming $p(x)$ is a good approximation of the stationary mispricing distribution, it immediately follows that $p(x)$ cannot remain unimodal whenever $Z(\Theta) < 0$, i.e. for $\Theta > \Theta_c \approx 0.797999$.

However, the quasi-stationary assumption is only approximate and the above prediction is not expected to be exact. In particular, when $\beta$/$\Theta$ increases, such an assumption is expected to be violated for two reasons: a) $\sigma_x$ increases, meaning that the dynamics of $x$ becomes more intense and ``blurs'' the distribution $p(M|x)$, lowering the feedback effect; b) as we show in the next section, the dynamics of $M$ slows down abruptly so that the separation of time scales becomes less and less warranted. Hence we expect that the above value of $\Theta_c$ is a lower bound to the exact value. 

Still, the above computation unveils the mathematical mechanism that leads to $p(x)$ becoming bimodal only for values of $\beta$ that are much larger than the ones that suffice for $p(M)$ to be bimodal. Indeed, the bimodality condition for $p(M)$, which can be calculated explicitly when $\kappa$ is small by expanding the normalisation $A(x)$, Eq.~\eqref{eq:E8}, of the conditional density $p(M|x)$ up to first order in $\kappa$:
\begin{align}
    A(x) 
    &= \left(\frac{1}{2}\right)^{\frac{2\beta}{\sigma_N^2\alpha\gamma}} \sqrt{\alpha\sigma_N^2 \pi} \, \, \mathrm{e}^{\frac{\beta^2 + \kappa^2 x^2}{\alpha \sigma_N^2}} E(\Theta)  + \mathcal{O} (\kappa^2).
\end{align}
Using this,  the conditional density reads
\begin{align}
    p(M|x) &\approx \frac{1}{A(x)}  \cosh (\gamma M)^{\frac{2\beta}{\sigma_N^2\alpha\gamma}} \, \mathrm{e}^{-\frac{M^2}{\alpha \sigma_N^2} + \frac{2\kappa}{\alpha \sigma_N^2} Mx} \\
    &= \frac{2^{\frac{2\beta}{\sigma_N^2\alpha\gamma}}}{\sqrt{\alpha\sigma_N^2 \pi} E(\Theta)} \mathrm{e}^{-\frac{\beta^2}{\alpha \sigma_N^2}} \mathrm{e}^{-\frac{(M-\kappa x)^2}{\alpha \sigma_N^2}} \cosh (\gamma M)^{\frac{2\beta}{\sigma_N^2\alpha\gamma}},
\end{align}
from which it follows that
\begin{align}
    p(M) &= \int_{-\infty}^{\infty} p(M|x) p(x) \,\mathrm{d}x \\
    &= \frac{1}{B} \frac{2^{\frac{2\beta}{\sigma_N^2\alpha\gamma}}}{\sqrt{\alpha\sigma_N^2 \pi} E(\Theta)} \mathrm{e}^{-\frac{\beta^2}{\alpha \sigma_N^2}} \cosh (\gamma M)^{\frac{2\beta}{\sigma_N^2\alpha\gamma}}
    \int_{-\infty}^{\infty} \mathrm{e}^{-\frac{(M-\kappa x)^2}{\alpha \sigma_N^2}}  \mathrm{e}^{-\frac{\kappa Z(\Theta)}{\sigma_x^2} x^2} \,\mathrm{d}x \\
    &=  \frac{2^{\frac{2\Theta^2}{\beta\gamma}}}{E(\Theta) e^{\Theta^2}} \sqrt{\frac{Z(\Theta)}{\pi (\sigma_x^2 \kappa + Z(\Theta) \alpha \sigma_N^2)}}  \, \cosh (\gamma M)^{\frac{2\Theta^2}{\beta\gamma}} \, \mathrm{e}^{-M^2 \frac{Z(\Theta)}{\sigma_x^2 \kappa + Z(\Theta) \alpha \sigma_N^2}},\label{eq:PM}
\end{align}

It is easy to see that $p(M)$ has an extremum at $M=0$ because the first derivative of the cosh, (minus) the sinh, evaluated at $M=0$ contributes a term equating zero and the derivative of the Gaussian contributes a term $\sim M$, which, of course, is also zero when $M=0$; in total: $p'(M=0) = 0$.
The second derivative (disregarding the normalisation) evaluated at $M=0$ is
\begin{equation}
    p''(M=0) \propto \left( \frac{2 \beta \gamma}{\alpha \sigma_N^2} - \frac{2 Z(\Theta)}{\sigma_x^2 \kappa + Z(\Theta) \alpha \sigma_N^2} \right) .
\end{equation}
The critical point $\Theta \equiv \Theta_c$ is the point where the mispricing distribution $p(x)$ becomes bimodal, and beyond which the Gaussian approximation for $p(x)$ breaks down entirely. Inserting this point into $p''(M=0)$ determines whether the trend distribution $p(M)$ is generally already bimodal when $p(x)$ just turns bimodal. Inserting the critical point yields
\begin{equation}
    p''(M=0)|_{\Theta = \Theta_c} \propto \frac{2\beta \gamma}{\alpha \sigma_N^2} >0
\end{equation}
because $Z(\Theta_c)=0$ and all parameters are positive. This shows that in this limit the trend distribution $p(M)$ always is bimodal \textit{before} the trend distribution becomes bimodal. In fact, $p(M)$ becomes bimodal as soon as $\beta > 1/\gamma$.

\subsubsection{Quasi-static Assumption Break-down}

In this section it will be shown why the quasi-static approximation from the previous section that works for small and moderate values of $\beta$ rapidly breaks down when $\beta$ is increased, such that no closed-form stationary distribution can be written down when $\beta$ becomes large -- approximately large enough to induce bimodality in $p(x)$.

The demonstration in this section is performed on the example of fixed $x=0$ for analytical tractability.
Note from the conditional FPE, Eq.~\eqref{eq:cond_FPE_M_given_X}, that $M$ moves like in an effective potential of $\frac{1}{2} M^2 - \frac{\beta}{\gamma} \ln (\cosh (\gamma M))$, corresponding to a force $\alpha M - \alpha\beta \tanh (\gamma M)$.
In the steady state one finds $M-\beta \tanh(\gamma M)=0$ for $M$. This equation shows a bifurcation: beyond the critical point $\beta\gamma=1$, the equation admits three solution, where the existence of three solutions corresponds to bimodality. Those solutions are $M=0$ and when $\beta\gamma>1$, as we have here, the other two solutions are $M\approx \pm \beta$ because the hyperbolic tangent will mostly be in its saturated regime, so at $\pm 1$, as, again, $\gamma M$ is large when $\beta\gamma$ is large.  For $\beta\gamma <1$, there is only one solution, $M=0$.

Plugging the bimodal case, where there are three solutions, back into the effective potential, yields an effective potential of zero when $M=0$ and an effective potential of
\begin{equation}
    \frac{1}{2} \beta^2 - \frac{\beta}{\gamma} \ln (\cosh (\pm \beta\gamma)) \approx \frac{1}{2} \beta^2 - \frac{\beta}{\gamma} |\beta\gamma| = -\frac{1}{2} \beta^2,
\end{equation}
for $M=\pm \beta$. The approximation is valid when $\ln (\cosh(x)) \approx |x|$.

From the Arrhenius law it then follows that the expected time to switch states from $M=-\beta$ to $M=+\beta$, i.e. the time $T_\times$ to cross the potential barrier, is of the order of \cite{hanggi1990reaction}
\begin{equation}
    T_\times \sim \frac{1}{\alpha} \mathrm{e}^{\frac{\beta^2}{2T}} = \frac{1}{\alpha} \mathrm{e}^{\Theta^2},
\end{equation}
where $T = \frac{\sigma_N^2 \alpha}{2}$ is the `temperature' parameter from statisitcal mechanics, which can be read off of the conditional FPE, Eq.~\eqref{eq:cond_FPE_M_given_X}.

This showcases that $M$ is no longer a fast variable when $\Theta$ is increased; as a matter of fact $M$ swiftly becomes very slow and the expected time to switch from, e.g.,  $+\beta$ to $-\beta$ diverges as $\beta$ is increased. This means that while the analytical distribution is still bimodal and symmetric in $M$, it will take exponentially longer to observe such a transition -- numerically this can no longer be observed.

Further, when $M$ suddenly becomes slow compared to $x$, the conditional FPE ansatz breaks down as $M$ no longer has the time to relax with respect to $x$. {\it In fine}, this is because the dominating time scale of $x$, $\kappa^{-1}$, is no longer much larger than $\alpha^{-1} e^{\Theta^2}$ -- in fact $\kappa^{-1}$ swiftly becomes much smaller, rendering the approximation invalid. Therefore, no closed-form solution can be derived in this way -- but the mechanism for bimodality can be revealed.

Interestingly, the break-down of the approximation and the onset of bimodality happen around the same parameter values because both the exponent of the Arrhenius law and the expression giving the value of $\kappa_{\rm eff}$ are functions of the very same combination of parameters $\Theta=\beta/\sqrt{\alpha \sigma_N^2}$.

\end{document}

%% file: preamble.tex
\usepackage{amsthm}
\usepackage{mathtools}
\usepackage{physics}
\usepackage{xcolor}
\usepackage{graphicx}
\usepackage[left=23mm,right=13mm,top=35mm,columnsep=15pt]{geometry} 
\usepackage{adjustbox}
\usepackage{placeins}
\usepackage[T1]{fontenc}
\usepackage{lipsum}
\usepackage{csquotes}